\documentclass[pra,aps,floatfix,superscriptaddress,showpacs,showkeys,amsmath,amssymb,nofootinbib,twocolumn,]{revtex4-1}
\usepackage{amsthm}
\usepackage{amsfonts}
\usepackage{amstext}
\usepackage{graphicx}
\usepackage{color}
\usepackage{enumitem}
\usepackage[dvipsnames]{xcolor}
\usepackage{tkz-kiviat,numprint} 
\usetikzlibrary{arrows}
\usetikzlibrary{plotmarks}
\usepackage{dcolumn}

\usetikzlibrary{shapes}
\usepackage{xspace}
\usepackage{mje}
\def\Bid{{\mathchoice {\rm {1\mskip-4.5mu l}} {\rm{1\mskip-4.5mu l}} {\rm {1\mskip-3.8mu l}} {\rm {1\mskip-4.3mu l}}}}

\newcommand{\A}{\ensuremath{{\hat{a}}\xspace}}

\newcommand{\Param}{\ensuremath{\boldsymbol{\Omega}}\xspace}
\newcommand{\Func}[1]{\ensuremath{\mathrm{W}_{#1}\left(\Param\right)}\xspace}

\newcommand{\OperBare}{\ensuremath{{\Delta}}\xspace}
\newcommand{\Oper}{\ensuremath{\OperBare\left(\Param\right)}\xspace}

\newcommand{\DO}{{{\rho}}}

\definecolor{mygreen}{RGB}{13,140,53}
\definecolor{myorange}{RGB}{220,60,30}

\newcommand{\QSERG}{Quantum Systems Engineering Research Group \& Department of Physics, Loughborough University, Leicestershire LE11 3TU, United Kingdom}
\newcommand{\TokyoTech}{Tokyo Institute of Technology, 2-12-1 Ookayama, Meguro-ku, Tokyo 152-8550, Japan}
\begin{document}


\title{Quantum phase space measurement and entanglement validation made easy}
\author{R. P. Rundle}
\affiliation{\QSERG}
\author{P.W. Mills}
\affiliation{\QSERG}
\author{Todd Tilma}
\affiliation{\TokyoTech}
\author{J. H. Samson}
\affiliation{\QSERG}
\author{M. J. Everitt}\email{m.j.everitt@physics.org}
\affiliation{\QSERG}
\date{\today}

\begin{abstract} 
It has recently been shown that it is possible to represent the complete quantum state of any system as a phase-space quasi-probability distribution (Wigner function) [Phys Rev Lett  117, 180401]. 
Such functions take the form of expectation values of an observable that has a direct analogy to displaced parity operators.  
In this work we give a procedure for the measurement of the  Wigner function that should be applicable to any quantum system. 
We have applied our procedure to IBM's \emph{Quantum Experience} five-qubit quantum processor to demonstrate that we can measure and generate the Wigner functions of two different Bell states as well as the five-qubit Greenberger-Horne-Zeilinger (GHZ) state.
As Wigner functions for spin systems are not unique, we define, compare, and contrast two distinct examples. 
We show how using these Wigner functions leads to an optimal method for quantum state analysis especially in the situation where specific characteristic features are of particular interest (such as for spin Schr\"odinger cat states). 
Furthermore we show that this analysis leads to  straightforward, and potentially very efficient, entanglement test and state characterisation methods.
\end{abstract}


\maketitle

\section{Introduction}
\begin{figure}[!b]
\begin{tikzpicture}[scale=1.0]
\pgftext[bottom,left]{
   \includegraphics[width=\linewidth,trim={210 150 130 100},clip = true]{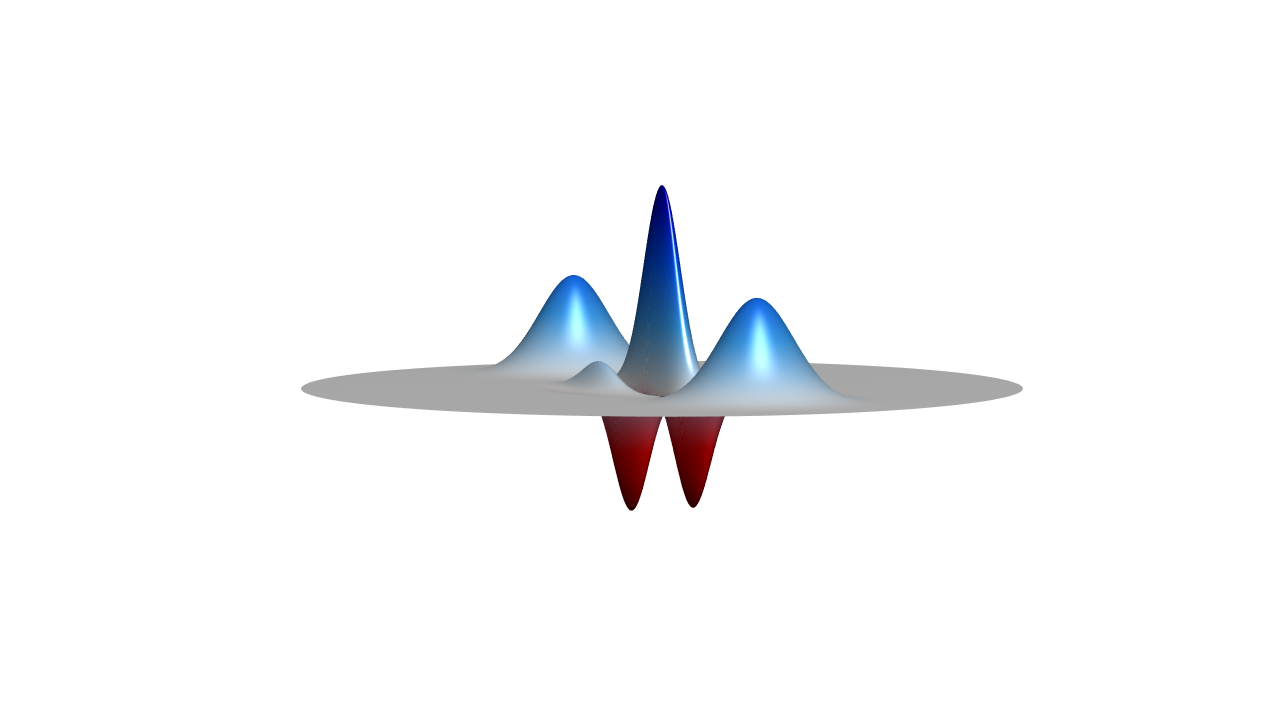}
};   
   \draw[thick] (5.4,2.5) -- (7,2.5) node[anchor=west] {\ket{\mathsf{alive}}};
   \draw[thick] (3.1,2.5) -- (2,2.5) node[anchor=east] {\ket{\mathsf{dead}}};
   \draw[thick] (4.4,3.5) -- (5.5,3.5) node[anchor=west] {\sf Quantum Interference};
   \draw[thick] (4.7,0.75) -- (5.5,0.75) node[anchor=west] {\sf Quantum Interference};
\end{tikzpicture}
    \caption{\label{fig:classic} The iconic textbook example of a Wigner function for a Schr\"odinger cat state. 
    The bell shapes represent the \emph{`alive'} and \emph{`dead'} possible states for the \emph{`cat'} and the oscillations between them indicate the quantum coherence between these states (\emph{i.\ e.\ }the classic ``both alive and dead'' statement). 
    A similar Wigner function without these interference terms would represent a state with a classical coin toss probability of being either ``alive'' or ``dead'' but not both. 
    The presence of the interference terms indicates that this Wigner function represents a state that is in both states (``alive and dead'') at the same time (a superposition).
    }
\end{figure}
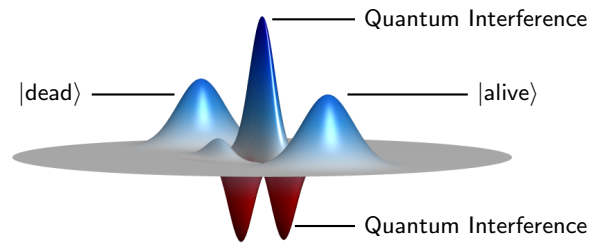

In 1932, Eugene Wigner, in an attempt to link the physics of many-particle systems (statistical physics) with quantum mechanics, defined a new way of describing the quantum state~\cite{PhysRev.40.749}. 
It took the form of a probability density function in position and momentum but, interestingly, it could take on negative values.
Now named after its creator, the Wigner function is usually presented in advanced quantum optics texts as an integral combining the notions of Fourier transformations and autocorrelations. 
The function rapidly established its usefulness when its ability to take on negative values enabled physicists to be able to visualise quantum correlations in ways that were not previously possible. 
This capability is most commonly seen in the superposition of two macroscopically distinct coherent states~\cite{PhysRevLett.10.277,PhysRev.131.2766,Carmichael}.
In~\Fig{fig:classic} we show an example of the Wigner function for such a superposition, the famous Schr\"odinger cat state.
Such a state is very similar to those presented in~\cite{Deleglise:2008gt} where it was demonstrated that non-classical states of light can be made.

Similar schemes to those used in~\cite{Deleglise:2008gt} for the direct reconstruction of the Wigner function for light have been in existence for some time (see, for example,~\cite{PhysRevLett.78.2547,PhysRevA.60.674,PhysRevLett.89.200402,PhysRevLett.87.050402,PhysRevA.70.053821}). 
These schemes all have the same feature that they, either implicitly or explicitly, rely on the fact that the Wigner function can be written as the expectation value of an appropriately normalized displaced parity operator or, equivalently, the expectation of parity for a displaced state~\cite{doi:10.1080/00107514.2010.509995}.
In quantum mechanics, parity is similar to the usual notion of point reflection in that it maps a co-ordinate to one of opposite sign, the difference being that the co-ordinate in quantum mechanics is an observable operator.  
What this means is that to reconstruct the Wigner function representation of the quantum state, all that is needed is a mechanism of displacing the quantum state and measuring its parity. 
Such operations are well established in the quantum optics community~\cite{PhysRevA.72.053818}. 
A similar procedure, designed and built around finite-dimensional systems, is however still lacking.

To address this lack of a mechanism for displacing the quantum state and measuring its parity for finite-dimensional quantum systems, we propose a phase-space formalism that allows for a full representation of a product Hilbert space and offers easily understandable visualizations. 
Focusing on the latter, the symmetric-subspace approach, for example the one presented in~\cite{PhysRevA.49.4101} where Wigner functions are constructed via a multi-pole expansion of spherical harmonics, is quite visually informative for harmonic-oscillator type systems~\cite{PhysRevA.6.2211,Kano1974} and those with spin-$1/2$ symmetry.
In more detail, it was Arecchi \textit{et.\ al.\ }\cite{PhysRevA.6.2211} that first derived spin-$1/2$ atomic coherent states described by continuous functions of Euler angles. 
These states satisfied the same mathematical properties as the Glauber-Sudarshan infinite-dimensional coherent states~\cite{Glauber1969,PhysRevLett.10.277} but offered discreteness and Bloch symmetry to the corresponding Hilbert space, thus allowing them to be used to describe an assembly of spin-$j$ particles. 
Soon after, Agarwal~\cite{Agarwal1981} rewrote the Wigner, $R$ and $P$ functions in terms of Arecchi's atomic coherent states, thus allowing for the study of various spin-$j$ systems under the Moyal quantization~\cite{Moyal1949}. 
These parametrizations allow for easy visualizations of various quantum systems via Dicke state mappings~\cite{PhysRev.93.99,1612-202X-4-12-009} to a multi-pole expansion of spherical harmonics, for example, but they do not allow for a full representation of a product Hilbert space. 
As such, all symmetric-subspace Wigner functions are limited insofar that they cannot correctly show entanglement or represent the set of states that lie outside of this subspace (which, for large numbers, is nearly all of the Hilbert space).

To address the issue of the full representation of a product Hilbert space, we propose that the the phase space needs to be parametrized by more generalized coherent states such as those derived by Nemoto~\cite{Nemoto2000} and Mathur \textit{et.\ al.\ }\cite{Mathur2002}. 
Such states can be used to construct characteristic functions beyond those written with atomic or three-level coherent states~\cite{Agarwal1981,Wolf6247,Luis-052112,Luis-495302,Klimov055303}. 
These characteristic functions~\cite{TilmaKae1}, by satisfying the Stratonovich-Weyl correspondence~\cite{Stratonovich56}, are informationally-complete \SU{N}-symmetric, spin-$j$ descriptions of finite-dimensional quantum states (``qudits'')~\cite{Shibata1976,PhysRevA.59.971,Braunstein-1004.5425,Klimov-1008.2920}.
This work is in contrast to that proposed by Wootters~\cite{WOOTTERS19871} and others for generating characteristic functions of $N$-dimensional discrete systems. 
There, the motivating mathematics are built around analyzing \emph{``systems having only a finite number of orthogonal states. The `phase space' for such a system is taken to be not continuous but discrete.''}~\cite{WOOTTERS19871}. 
The phase space generated by such generalized coherent states is continuous in its parametrization (see~\cite{Boya2,UandCPN}), allows for Wigner functions to be generated by the methodology given in~\cite{1601.07772} (the expectation value of an appropriately normalized displaced general parity operator), can completely represent product Hilbert spaces of qudits (thus producing phase space signatures of entanglement) and gives a method for visualizing said functions that is equivalent to that done for symmetric subspace representations, which we now discuss in more detail.

\section{Background}
While it has been known for a long time that parity displacement could be done for continuous systems~\cite{Klauder-Sudarshan,Glauber1969}, following much work on the use of Wigner functions of discrete systems~\cite{Wigner1984,Agarwal1981,PhysRevA.59.971,Braunstein-1004.5425,Klimov-1008.2920,Luis-052112,Luis-495302,Klimov055303,TilmaKae1,Weyl1927,Moyal1949,Varilly:1989gs,WOOTTERS19871,PhysRevA.53.2998,VOURDAS1997367,PhysRevA.65.062309,PhysRevA.70.062101,Wolf6247}, it has only recently been proposed that any quantum system's Wigner function can be written as the expectation value of a displaced and/or rotated generalized parity operator~\cite{1601.07772}. 
Mathematically this can be expressed as
\ba
\label{rDO0}
W_\rho(\Param)&=&\EX{U(\Param) \Pi U^\dag(\Param) }_\rho \nonumber \\
&=&\Trace{\DO  \left\{U(\Param) \Pi U^\dag(\Param) \right\}}
\ea
where $W$ is the Wigner function and $\Param$ is the set of parameters over which displacement or rotations are defined (typically this would be position and momentum); $\DO$ is the density matrix; $U(\Param)$ is a general displacement/rotation operator, or collection of operators; and $\Pi$'s definition is motivated by the usual parity operator.
The conventional Wigner function in position and momentum space is obtained if $U$ is set to the displacement operator that defines coherent states, \ket{\alpha}, from the vacuum state, \ket{0}, according to $D(\alpha)\ket{0}=\ket{\alpha}$ and the operator $\Pi$ is defined to be twice the usual phase space parity operator so that $\Pi \ket{\alpha}=2\ket{-\alpha}$~\cite{PhysRevA.50.4488}.

For a given system the choice of $U(\Param)$ and $\Pi$ is not unique but in~\cite{1601.07772} it was stipulated that a distribution $W_{{\rho}}(\Omega)$ over a phase space defined by the parameters $\Omega$ is a Wigner function of ${\rho}$ if there exists a kernel ${\Delta}(\Omega)$ (which we show can be written as a similarity transform, with respect to a ``displacement'', of a parity-like operator, i.e. ${\Delta}(\Omega)=U(\Param) \Pi U^\dag(\Param)$ - and the Wigner function is the expectation value of this similarity-transformed operator) satisfying the following restricted version of the Stratonovich-Weyl correspondence (reproduced verbatim from~\cite{1601.07772}): 
\begin{enumerate}[label=\sffamily \footnotesize \upshape S-W.\arabic*]
\item\label{D1} The mappings $\Func{\DO}=\Trace{\DO \, \Oper}$ and $\DO = \int_{\Param} \Func{\DO} \Oper \ud \Param$ exist and are informationally complete. Simply put, we can fully reconstruct $\DO$ from $\Func{\DO}$ and vice versa\footnote{For the inverse condition, an intermediate linear transform may be necessary.}.
\item\label{D2} $\Func{\DO}$ is always real valued which means that $\Oper$ must be Hermitian. 
\item\label{D3}  $\Func{\DO}$ is ``standardized'' so that the definite integral over all space $\int_{\Param} \Func{\DO} \ud \Param = \Tr{\DO}$  exists and $\int_{\Param} \Oper \ud \Param =\Bid$.
\item\label{D4} Unique to Wigner functions, $\Func{\DO}$ is self-conjugate; the definite integral $\int_{\Param} \Func{\DO'}\Func{\DO''} \ud \Param= \Trace{\DO' \DO''} $ exists. 
This is a restriction of the usual Stratonovich-Weyl correspondence. 
\item\label{D5} Covariance:
Mathematically, any Wigner function generated by ``rotated'' operators ${\Delta}(\Param^{\prime})$ (by some unitary transformation $V$) must be equivalent to ``rotated'' Wigner functions generated from the original operator (${\Delta}(\Param^{\prime}) \equiv V \Oper V^{\dagger}$) - \textit{i.\ e.\ }if $\DO$ is invariant under global unitary operations then so is $\Func{\DO}$.
\end{enumerate}
If we define $U(\Param)$ as an element of a Special Unitary (SU) group that acts as a displacement or rotation and $\Pi$ as an appropriately normalised identity plus a traceless diagonal matrix (i.e.\ an element of the Cartan sub-algebra of the appropriate group) then, from~\cite{1601.07772},~\Eq{rDO0} is sufficient to generate Wigner functions for any finite-dimensional, continuous-variable, quantum system. 
We note that beyond satisfying the Stratonovich-Weyl correspondence, we have yet to fully determine the level to which this definition is constrained. 
Because $\Pi$ performs the same role as parity does in the standard Wigner function, we refer to it as an \emph{extended parity}.

\section{The Scheme}
In this work we present a procedure for the measurement and reconstruction of the quantum state for a series of qubits from two different Wigner functions that both satisfy the above restricted Stratonovich-Weyl correspondence. 
We start by considering a Wigner function where the extended parity operator is defined with respect to the underlying group structure of the total system. 
We then proceed to investigate another Wigner function, whose kernel comprises a tensor product of one-qubit kernels, which is arguably a more natural way of looking at composite quantum systems.
In both cases we apply our procedure to IBM's \emph{Quantum Experience} five-qubit quantum processor to demonstrate that we can measure and reconstruct the Wigner functions of two different Bell states and the five-qubit Greenberger-Horne-Zeilinger (GHZ) state.  

While Wigner functions can be considered to be expectation values of displaced extended parity operators, this view does not necessarily lead to the best way to practically determine the Wigner function.   
As previously discussed, displacing the extended parity operator and taking its expectation value should be the same as displacing/rotating the state, \textit{i.\ e.\ }creating a new ``state''
\bel{rDO} 
\tilde\DO(\Param)=U^\dag(\Param) \DO U(\Param),
\ee 
and calculating the expectation value of the unshifted extended parity operator.
\bel{rDO2}
\EX{ \Pi  }_{\tilde\DO(\Param)}=\Trace{\tilde\DO(\Param) \Pi}.
\ee
Mathematically this is equivalent to our original expression for the Wigner function (\Eq{rDO0}) as trace is invariant under cyclic permutations of its arguments.
Furthermore, it is possible, and in some cases (such as with the IBM \emph{Quantum Experience}) easier, to make $\tilde\DO(\Param)$ by performing local rotations on each qubit rather than displacing $\Pi$.

In the ideal case, the extended parity $\Pi$ shown in~\Eq{rDO2} will be directly measurable, allowing for reconstruction of the quantum state via its Wigner function without any intermediate steps being needed. 
Even if it is not possible to measure the extended parity directly, such as with the IBM \emph{Quantum Experience}, there is a simple alternative. 
Note that $\Pi$, as introduced in~\cite{1601.07772}, is always a diagonal operator in the computational basis.
The Wigner function is then easy to calculate according to
\bel{sum}
W(\Param)=\sum_n{\tilde\DO_{nn}(\Param)}\Pi_{nn}.
\ee
To determine the Wigner function we are only required to measure the probability of the rotated system occupying each state of the computational basis.

For a set of qubits the rotation of the system can be intuitively defined in terms of rotation operators acting on each of the system's constituent parts. 
Explicitly, we can define a total rotation operator for $N$ qubits as 
\be
{\mathbb{U}}_N=\bigotimes_i^N {U}_i(\theta_i,\varphi_i,\Phi_i)
\ee 
where $ {U}_i(\theta_i,\varphi_i,\Phi_i)=e^{\ui {\sigma}_{z_{i}} \varphi_i}e^{\ui {\sigma}_{y_{{i}}} \theta_i}e^{\ui{\sigma}_{z_{i}} \Phi_i}$ is the \SU{2} rotation operator for each qubit in terms of the Euler angles $\Param_i=(\theta_i,\varphi_i,\Phi_i)$.  
In the following sections we discuss the Wigner functions defined through two different possible choices of $\Pi$.

\section{A Spin Wigner function with \SU{\cdot} Extended Parity}
In this section we define and explore a Wigner function for $N$ qubits where the extended parity operator reflects the underlying group structure of the total system.  
Here, extended parity is motivated by the idea of doing what amounts to a global $\pi$ rotation on the hypersphere of the underlying \SU{2^{[N]}} coherent state representation.
This is achieved by defining our extended parity operator $\Pi_{\SU{2^{[N]}}}$ as a $2^N \times 2^N$ diagonal matrix whose first element\footnote{This particular representation of extended parity is a rotation of the extended parity operator given in~\cite{1601.07772} that we have taken in order to keep within the conventions of the experimental physics and quantum information communities. As with the extended parity operator given in~\cite{1601.07772}, ours is still a linear function of the identity plus the Cartan sub-algebra of the selfsame SU group.} is $2^{-N}\left[1+(2^N-1)\sqrt{2^N+1}\right]$ and whose remaining diagonal elements are $2^{-N}\left[1-\sqrt{2^N+1}\right]$. For example, 
\bel{pione}
\Pi_{\SU{2^{[1]}}}=\Half\begin{pmatrix}
1+\sqrt{3} & 0   \\ 
0&1-\sqrt{3}  \\ 
\end{pmatrix}
=
\Half[\Bid+\sqrt{3}\sigma_z]
\ee
for one qubit and
\ba
\label{parity}
\Pi_{\SU{2^{[2]}}}&=&\frac{1}{4}
\begin{pmatrix}
1+3\sqrt{5} & 0 & 0 & 0 \\ 
0&1-\sqrt{5} & 0 & 0 \\ 
0&0&1-\sqrt{5} & 0  \\ 
 0 & 0 & 0 &1-\sqrt{5}\\ 
\end{pmatrix} \\ 
&=&
\frac{1}{4}[\Bid \otimes \Bid + \sqrt{5} \, \Bid \otimes \sigma_z + \sqrt{5} \, \sigma_z \otimes \Bid + \sqrt{5} \sigma_z \otimes \sigma_z] \nonumber
\ea
for two qubits in the computational basis.

Combining this definition of extended parity with the composite rotation operator, $\mathbb{U}_N$  we obtain the kernel
\be 
\OperBare_{\SU{2^{[N]}}}(\{\theta_i,\varphi_i\}) = 
	\mathbb{U}_{N}
	\hat{\Pi}_{\SU{2^{[N]}}}
	\mathbb{U}_N^\dag
\ee
that satisfies the restricted Stratonovich-Weyl correspondence given in the introduction.
We note that the $\Phi_i$'s make no contribution as $\Pi_{\SU{2^{[N]}}}$ commutes with $\sigma_{z_i}$. 
This kernel defines our $\SU{2^{[N]}}$, extended parity-based, Wigner function according to
\bel{Wparity}
W_{\SU{2^{[N]}}}(\{\theta_i,\varphi_i\})= \Trace{\rho \, \mathbb{U}_N
	\hat{\Pi}_{\SU{2^{[N]}}}
	\mathbb{U}_N^\dag}.
\ee

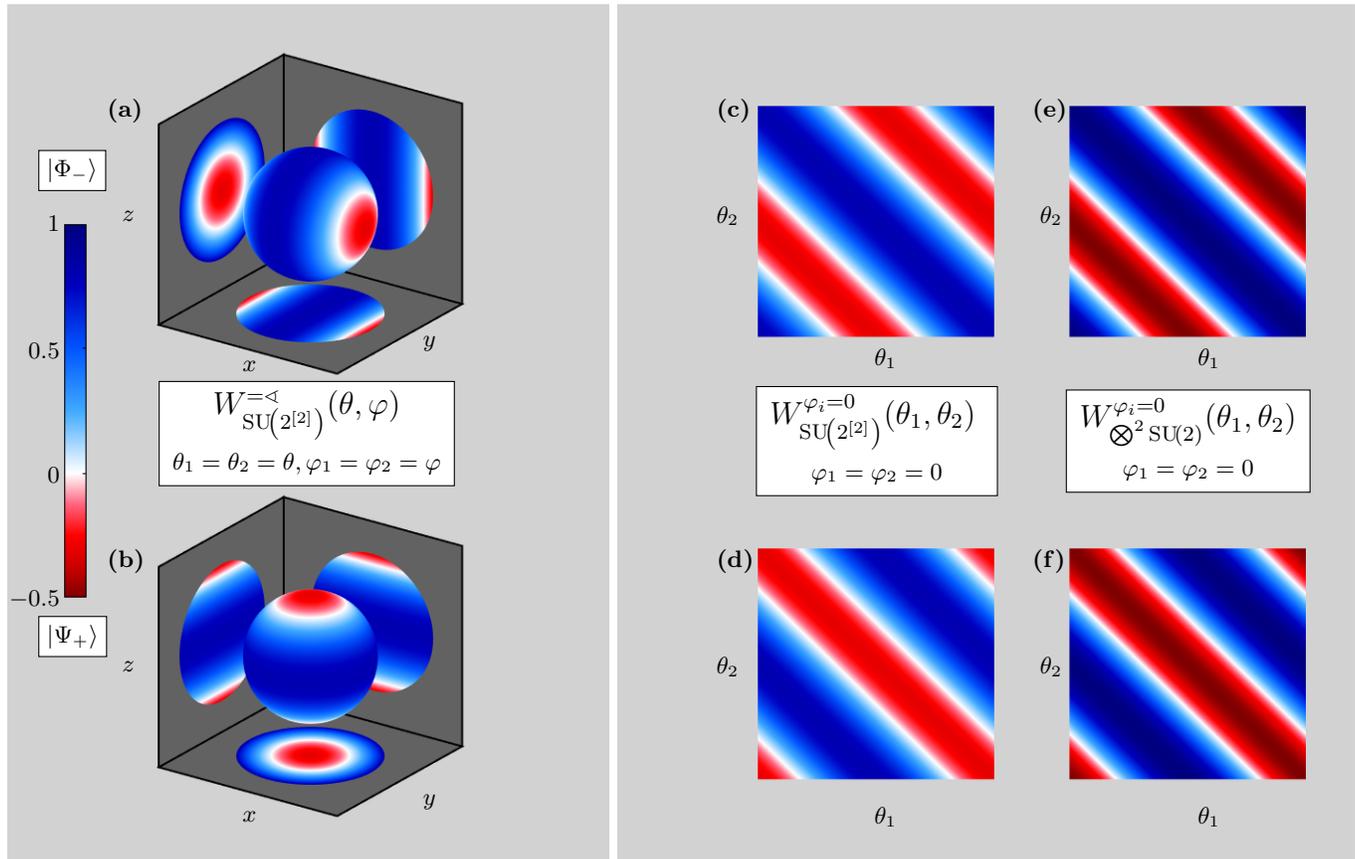
\begin{figure*}[!tb]

\begin{tikzpicture}[scale=1.0]
\pgftext[bottom,left]{%
    \includegraphics[width=\linewidth,trim=40 50 100 20,clip=true]{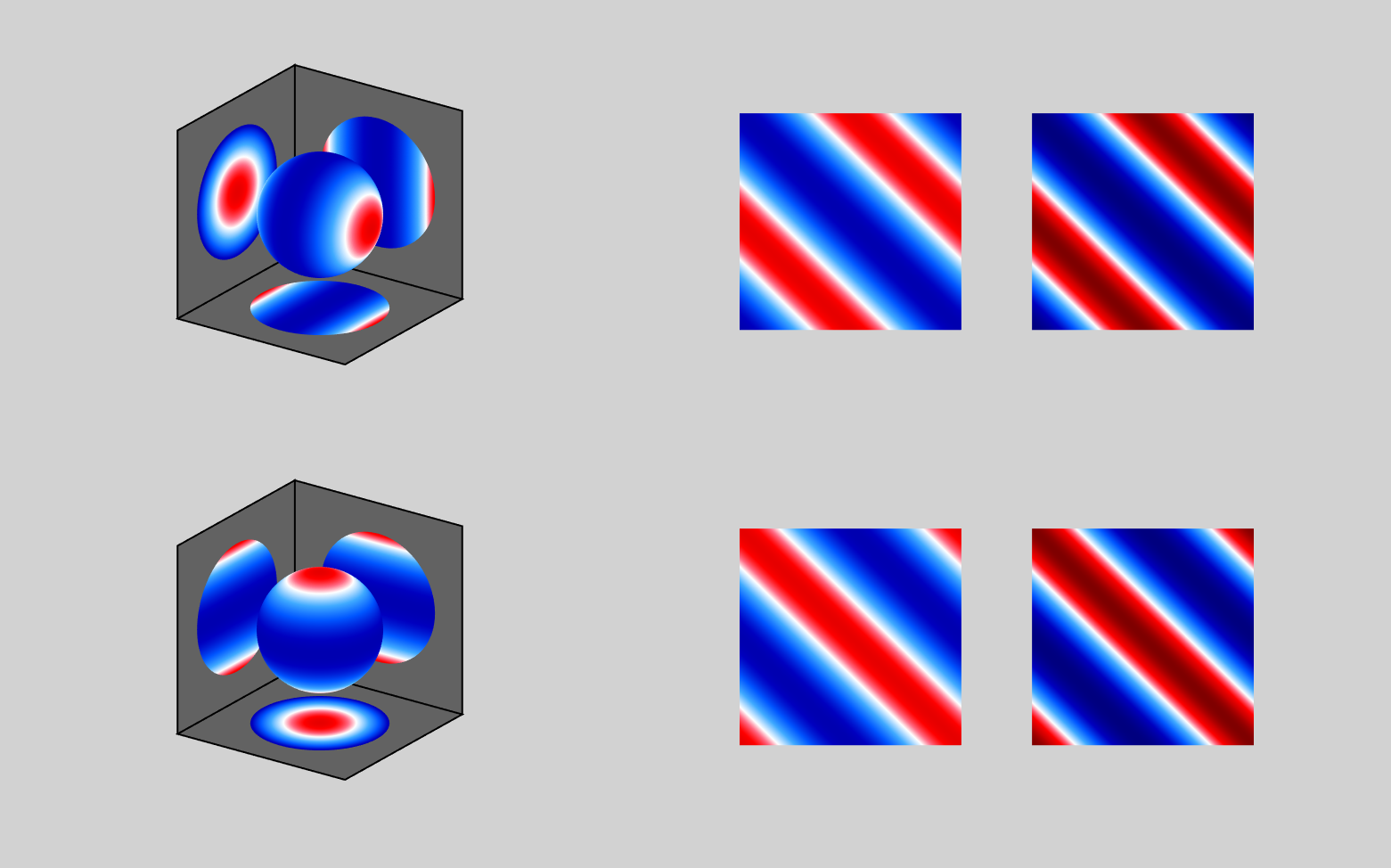}%
}%

\def \lH {2.2}
\def \blH {1}
\node[anchor=west] at (1.2,12.2-\lH) {{\bf (a)}};
\node[anchor=west] at (1.2,5-\blH) {{\bf (b)}};
\node[anchor=west] at (9.3,12.2-\lH) {{\bf (c)}};
\node[anchor=west] at (9.3,5-\blH) {{\bf (d)}};
\node[anchor=west] at (13.5,12.2-\lH) {{\bf (e)}};
\node[anchor=west] at (13.5,5-\blH) {{\bf (f)}};

\def \cH {6}
\node at (0.9,\cH){   \includegraphics[height=5cm]{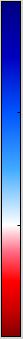}%
};
\node[align=right,anchor=east] at (0.8,\cH-2.5) {$-0.5$};
\node[align=right,anchor=east] at (0.8,\cH+0.8) {$0.5$};
\node[align=right,anchor=east] at (0.8,\cH+2.5) {$1$};
\node[align=right,anchor=east] at (0.8,\cH-0.85) {$0$};
\def \axH {1.6}
\node[anchor=west] at (3,8.2-\axH) {$x$};
\node[anchor=west] at (3,0.6) {$x$};
\node[anchor=west] at (5.4,8.5-\axH) {$y$};
\node[anchor=west] at (5.4,0.8) {$y$};
\node[anchor=west] at (1.4,10.2-\axH) {$z$};
\node[anchor=west] at (1.4,2.6) {$z$};

\node[anchor=west,rectangle,draw, fill=white] at (0.4,9.2) {$\ket{\Phi_-}$};
\node[anchor=west,rectangle,draw, fill=white] at (0.4,3) {$\ket{\Psi_+}$};

\def \H {5.7}

\node[anchor=west,rectangle,draw, fill=white] at (2,\H) {\begin{tabular}{c}
{\large $W_{\SU{2^{[2]}}}^{=\sphericalangle}(\theta, \varphi)$}\vspace*{4pt}\\
$\theta_1=\theta_2=\theta, \varphi_1 =\varphi_2=\varphi$
\end{tabular}
};
\node[anchor=west,rectangle,draw, fill=white] at (9.94,\H-0.1) {\begin{tabular}{c}
{\large $W^{\varphi_i=0}_{\SU{2^{[2]}}}(\theta_1, \theta_2)$} \vspace*{4pt}\\
$\varphi_1 =\varphi_2=0$
\end{tabular}
};
\node[anchor=west,rectangle,draw, fill=white] at (14.06,\H-0.1) {\begin{tabular}{c}
{\large $W^{\varphi_i=0}_{\bigotimes^2\SU{2}}(\theta_1, \theta_2)$}\vspace*{4pt} \\
$\varphi_1 =\varphi_2=0$
\end{tabular}};

\node[anchor=west] at (11.4,1.6-\blH) {$\theta_1$};
\node[anchor=west] at (9.3,3.6-\blH) {$\theta_2$};
\node[anchor=west] at (11.4,8.9-\lH) {$\theta_1$};
\node[anchor=west] at (9.3,10.8-\lH) {$\theta_2$};
\def \dX {4.3}
\node[anchor=west] at (11.4+\dX,1.6-\blH) {$\theta_1$};
\node[anchor=west] at (9.3+\dX,3.6-\blH) {$\theta_2$};
\node[anchor=west] at (11.4+\dX,8.9-\lH) {$\theta_1$};
\node[anchor=west] at (9.3+\dX,10.8-\lH) {$\theta_2$};

\fill[white] (8, 0) rectangle (8.1,13.65);

\end{tikzpicture}
    \caption{    \label{fig:example} \textbf{(a-d)} Slices from the four-dimensional Wigner function $W_{\SU{2^{[2]}}}(\theta_1, \varphi_1, \theta_2, \varphi_2)$ of two qubits for two different, maximally entangled, Bell states $\ket{\Phi_-}=(\ket{0}_1\ket{0}_2-\ket{1}_1\ket{1}_2)/\sqrt{2}$ and $\ket{\Psi_+}=(\ket{0}_1\ket{1}_2+\ket{1}_1\ket{0}_2)/\sqrt{2}$. 
    The three-dimensional plots \textbf{(a,b)} show $W_{\SU{2^{[2]}}}^{=\sphericalangle}(\theta, \varphi)$, the slice where $\theta=\theta_1=\theta_2$ and $\varphi=\varphi_1=\varphi_2$. 
    The two-dimensional plots \textbf{(c,d)} of $\theta_1$ versus $\theta_2$ show $W^{\varphi_i=0}_{\SU{2^{[2]}}}(\theta_1, \theta_2)$, the slice where $\varphi_1=\varphi_2=0$. 
    We recommend that the reader see the supplementary material which expands on these figures and shows animations of the Deutsch-Jozsa algorithm~\cite{DeutschJoza} and the creation of all four Bell states (in the animations, for example, it becomes clear that the Wigner functions for the Bell states [or for that matter, any maximally entangled two qubit state] are simply rotations of the same function in four-dimensional space). 
    Later in this work we will present experimental reconstructions of the $\theta_1$ versus $\theta_2$ plots. 
    In understanding the form of these plots we note that the $\ket{\Psi_+}$ state is one with total spin-angular momentum $\hbar$ but zero total $z$ spin-angular momentum. 
    We thus expect to see the observed ring-like symmetry in $W_{\SU{2^{[N]}}}^{=\sphericalangle}(\theta, \varphi)$ for $\ket{\Psi_+}$ (the symmetry of $\ket{\Phi_-}$ follows from $\ket{\Psi_+}$ as they are rotations of each other in four-dimensional space). 
    This state is also an angular-momentum analogue of a photon number (Fock) state which shares a similar symmetry in its Wigner function~\cite{PhysRevA.70.053821,PhysRevLett.87.050402,PhysRevLett.89.200402}. 
    In \textbf{(e,f)} we show $W^{\varphi_i=0}_{\bigotimes^2\SU{2}}(\theta_1, \theta_2)$ created using the alternative extended parity operator $\Pi_{\bigotimes^2\SU{2}}$ as discussed in section~\ref{TPsec}. 
    The availability of more than one extended parity operator, which  produces Wigner functions with qualitatively very similar features, opens up possible alternative paths for direct phase space reconstruction (note we have also included an animation of $W_{\bigotimes^2\SU{2}}$ for the creation of the Bell states in the supplementary material).}
\end{figure*}

Let us now consider the specific case of the Wigner function $W_{\SU{2^{[N]}}}$ for two qubits.
Each qubit brings with it two degrees of freedom, expressed in terms of Euler angles $\Param=(\theta_1,\varphi_1,\theta_2,\varphi_2)$, thus the associated Wigner function takes the form of a four-dimensional pseudo-probability distribution $W_{\SU{2^{[2]}}}(\theta_1,\varphi_1,\theta_2,\varphi_2)$. 
Four-dimensional functions are not easy to visualise, but we can take slices of the function in order to gain an appreciation of it as a whole. 
In~\Fig{fig:example}(a-d) we show some example Wigner function slices for two Bell states. 
Specifically~\Fig{fig:example}(a,b) shows the equal angle (``$=\sphericalangle$'') slice $W_{\SU{2^{[2]}}}^{=\sphericalangle}(\theta,\varphi)=W_{\SU{2^{[2]}}}(\theta,\varphi,\theta,\varphi)$ while~\Fig{fig:example}(c,d) shows the slice $W^{\varphi_i=0}_{\SU{2^{[2]}}}(\theta_1,\theta_2)=W_{\SU{2^{[2]}}}(\theta_1,0,\theta_2,0)$. Note that~\Fig{fig:example}(e,f) will be discussed in section~\ref{TPsec}.

In order to demonstrate that this function is indeed easy to construct we have taken advantage of IBM's \emph{Quantum Experience} project.
The project makes available through the Internet a five-qubit processor, initially based on a simple ``star'' topology\footnote{Before the early 2017 update by IBM}: a central qubit is coupled to four other qubits. 
The machine has already been used to produce interesting results~\cite{1605.05709,1605.04220}.
Here we use it to measure and reconstruct the Wigner functions for the two Bell states $\ket{\Phi_+}$ and $\ket{\Psi_-}$ as presented in~\Fig{fig:example}. 
In this work, we are limited by the operations that IBM has made available to the user, operations that naturally focus on quantum computing applications. 
Nevertheless, following~\Eq{rDO}, we are able to produce $\tilde\DO(\theta_1,\varphi_1,\theta_2,\varphi_2)$ using rotations generated by combinations of gate operations and readout state populations of $\tilde\DO_{nn}(\theta_1,\varphi_1,\theta_2,\varphi_2)$ via the standard output of the IBM processor. 
We then use~\Eq{sum} and~\Eq{parity} to reconstruct the Wigner function,~\Eq{Wparity}. 

\begin{figure}[!t]
\begin{tikzpicture}[scale=1.0]

\node[inner sep=0pt] at (0,0)
    {\includegraphics[width=\linewidth]{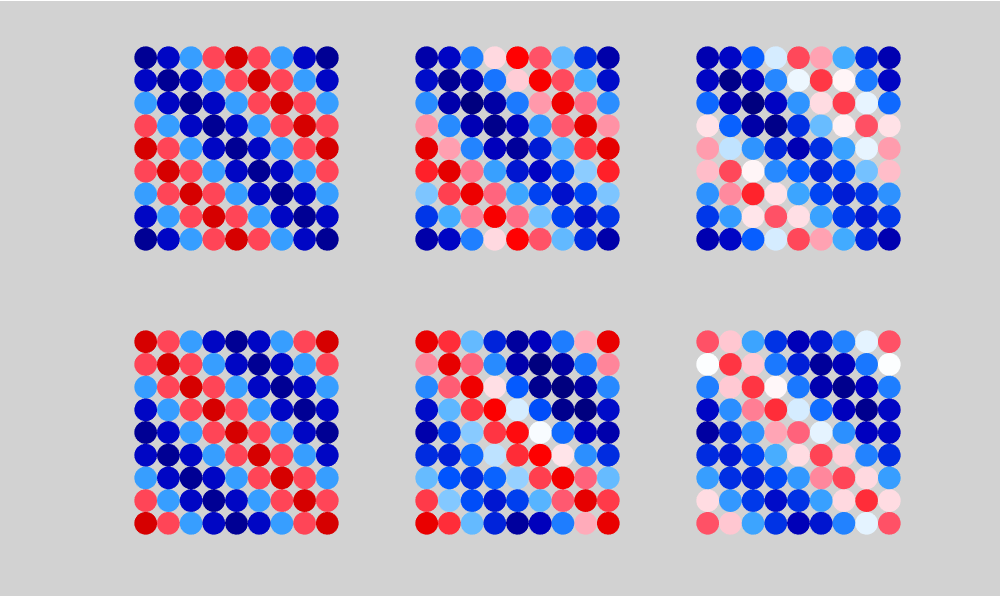}};
\node[inner sep=0pt] at (0,-3.2)
    {\includegraphics[width=\linewidth,trim=0 150 0 60,clip=true]{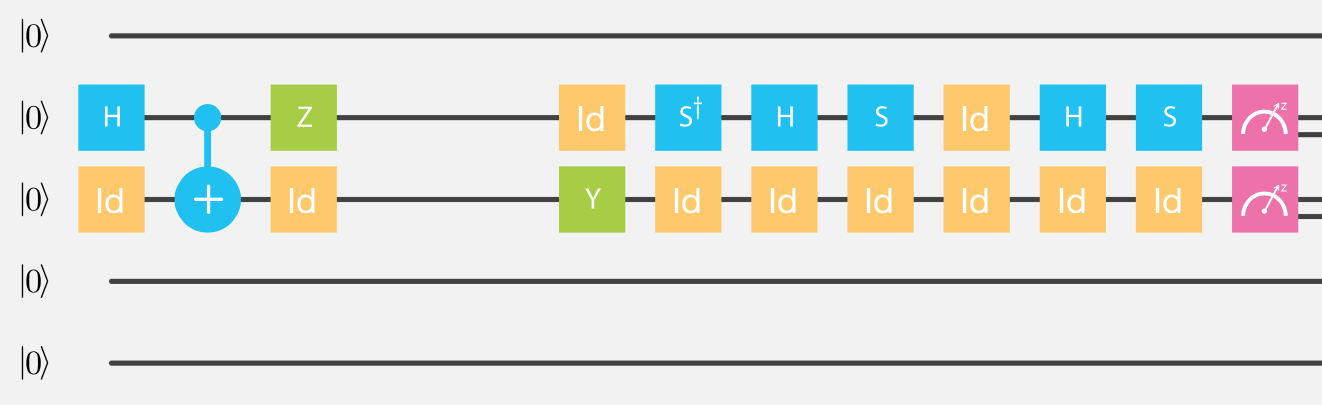}};
\node[anchor=west,rectangle,draw, fill=white] at (-2.9,2.6) {\sf Theory};
\node[anchor=west,rectangle,draw, fill=white] at (-0.66,2.6) {\sf Simulated};
\node[anchor=west,rectangle,draw, fill=white] at (1.6,2.6) {\sf Experimental};

\def \Xs {3.3}; 
\def \Ys {-4.7}; 
\node at (0.9+\Xs,4+\Ys){   \includegraphics[height=3cm]{colourbar}%
};
\node[align=right,anchor=east] at (0.85+\Xs,2.5+\Ys) {\footnotesize$-0.5$};
\node[align=right,anchor=east] at (0.85+\Xs,4.5+\Ys) {\footnotesize$0.5$};
\node[align=right,anchor=east] at (0.85+\Xs,5.45+\Ys) {\footnotesize$1$};
\node[align=right,anchor=east] at (0.85+\Xs,3.5+\Ys) {\footnotesize$0$};

\node[anchor=west,rectangle,draw, fill=white] at (-4.3,1.8) {$\ket{\Phi_-}$};
\node[anchor=west,rectangle,draw, fill=white] at (-4.3,-0.5) {$\ket{\Psi_+}$};

\node[anchor=center] at (-3.0,-4.2) {$\underbrace{\hspace{45pt}}_{\text{\footnotesize \sf Gates for \ket{\Phi_-}}}$};
\node[anchor=center] at (1.4,-4.2) {$\underbrace{\hspace{120pt}}_{\text{\footnotesize \sf Gates for performing $\theta$ rotations}}$};
\node[anchor=center] at (4.05,-4.2) {$\underbrace{\hspace{10pt}}_{\text{\footnotesize \sf Readout}}$};

\def \dX {0.1}
\node[anchor=west] at (0,0.2) {$\theta_1$};
\node[anchor=west] at (-2.5,0.2) {$\theta_1$};
\node[anchor=west] at (2.5,0.2) {$\theta_1$};
\node[anchor=west] at (0,-2.2) {$\theta_1$};
\node[anchor=west] at (-2.5,-2.2) {$\theta_1$};
\node[anchor=west] at (2.5,-2.2) {$\theta_1$};
\node[anchor=west] at (-3.6-\dX,-1.1) {$\theta_2$};
\node[anchor=west] at (-1.15-\dX,-1.1) {$\theta_2$};
\node[anchor=west] at (1.3-\dX,-1.1) {$\theta_2$};
\node[anchor=west] at (-3.6-\dX,1.3) {$\theta_2$};
\node[anchor=west] at (-1.15-\dX,1.3) {$\theta_2$};
\node[anchor=west] at (1.3-\dX,1.3) {$\theta_2$};

\end{tikzpicture}
    \caption{\label{fig:IBM2qubit} Plots of the spin Wigner function for the two Bell states $\ket{\Phi_+}$ and $\ket{\Psi_-}$. 
    We plot $\theta_1$ versus $\theta_2$ for the $W_{\SU{2^{[2]}}}^{\varphi_i=0}$ slice of the Wigner function for two qubits; making use of the periodicity of the function at the edges of each plot for computational efficiency. 
    We have included for comparison ideal theoretical values, numerical results using IBM's built in simulator, and real experimental data from IBM's quantum processor. 
    The quantum circuit presented above is a screenshot taken directly from IBM's \emph{Quantum Experience} web interface.
    It provides an example of the measurement protocol we used to obtain the diagonal elements of the rotated density matrix $\tilde\DO_{nn}(\theta_1,\varphi_1,\theta_2,\varphi_2)$. 
    The theoretical, simulated, and experimental data are all in very good agreement with each other. 
    Slight differences exist due to imperfect implementation of needed rotations due to different gate operations having different levels of noise (decoherence).
    It should be straightforward to replace the ``Gates for performing $\theta$ rotations'' with generalized rotation operators on each qubit.
    Furthermore, if measurement of the extended parity operator ($\Pi$) were available, direct observation of the quantum state would be reduced to a two-stage process of rotate and measure. 
    We believe such a protocol, because it would need fewer gate operations, would result in better agreement between theory and experiment than that seen in this figure. Note that in order to have good colour graduation in the transition from positive to negative values there is some color clipping for the very strong blue points.}
\end{figure}

In~\Fig{fig:IBM2qubit} we plot the Wigner function $W_{\SU{2^{[2]}}}^{\varphi_i=0}(\theta_{1},\theta_{2})$ slices  comparing the ideal theoretical values of~\Fig{fig:example}(c,d), values generated by IBM's built in simulator (that models environmental effects), and real experimental data.  
The calibration data pertaining to the experiments is provided in Table~\ref{table}.
In principle, to fully reconstruct the state requires us to measure the same number of points as needed to reconstruct the density matrix.
In~\Fig{fig:IBM2qubit} we have actually measured more, and different, points than would be needed to fully reconstruct the state.
This was done to demonstrate the ability to generate the Wigner function using a raster scan approach as this makes clear the straightforward nature of our measurement method. 
Due to finite computational resources, and the need to do rotations as outlined above, we are limited in our resolution. 
Nevertheless, we find good agreement between theory, simulation, and experimental data, demonstrating that our tomographic process is clearly able to distinguish between the two Bell states. 
\newcolumntype{d}[1]{D{.}{.}{#1} }
\begin{table}
\begin{tabular}{|l | d{-1} d{-1} | d{-1}d{-1}d{-1}d{-1}d{-1}|}
\cline{2-8}
\multicolumn{1}{ c|}{}&  \multicolumn{2}{ c|}{Bell} & \multicolumn{5}{ c|}{GHZ}\\ \hline
     qubit                                            &     \mathrm{1}  &  \mathrm{2}   &       \mathrm{0} &  \mathrm{1}   &  \mathrm{2} &  \mathrm{3} &  \mathrm{3}\\ \hline  \hline
$T_1 (\mathrm{\mu s})$                   &   85.8                &   75.1          &   58.9                &   87.1              &    74.7           & 74.8  & 65.5\\
$T_2 (\mathrm{\mu s})$                    &   109.6                &   58.8          &   74.8                &   142.2            &        59.2      & 53.2 & 48.4\\
$\epsilon_g (\times 10^{-2})$          &   0.15                &   0.2          &    0.29                &    0.2               & 0.23             & 0.23 & 0.89\\
$\epsilon_r (\times 10^{-2})$           &   4.6                &   4.3         &   4.6                   &   4.2                &  3.6              & 3.6    & 5.7\\
$\epsilon_g^{i2} (\times 10^{-2})$  &   3.19                &          & 5.21                    &  3.31               &                  &  3.18    & 6.55 \\
\hline
\end{tabular}
\caption{Calibration data for the experimental results contained within this paper. 
Data for the Bell state and GHZ Wigner functions were taken on 16\Th and 17\Th June 2016 when the fridge temperature was 18.25\,mK and 17.916\,mK respectively. 
$T_1$ and $T_2$ are the usual relaxation times, $\epsilon_g$ is the gate error,  $\epsilon_r$ is the readout error and $\epsilon_g^{i2}$ is the C-NOT gate error between the qubit listed and qubit 2 (which is the target qubit for the C-NOT operation).
\label{table}}
\end{table}

\begin{figure}[!t]
\begin{tikzpicture}[scale=1.0]

\node[inner sep=0pt] at (0,0)
    {\includegraphics[width=\linewidth]{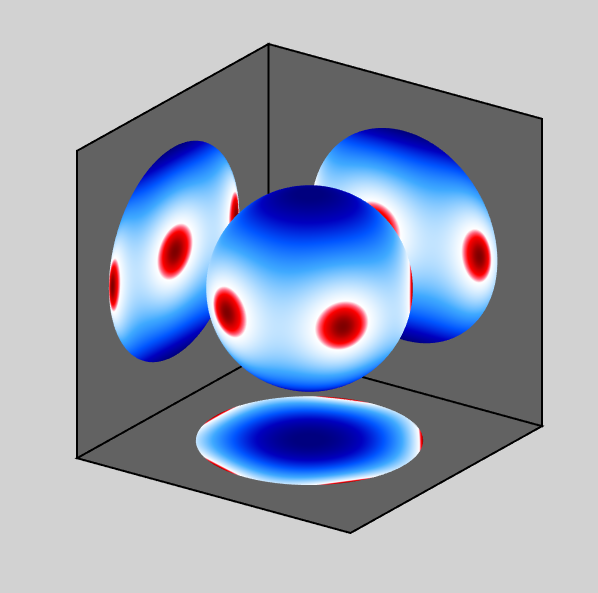}};
\node[inner sep=0pt] at (0,-5.2)
    {\includegraphics[width=\linewidth]{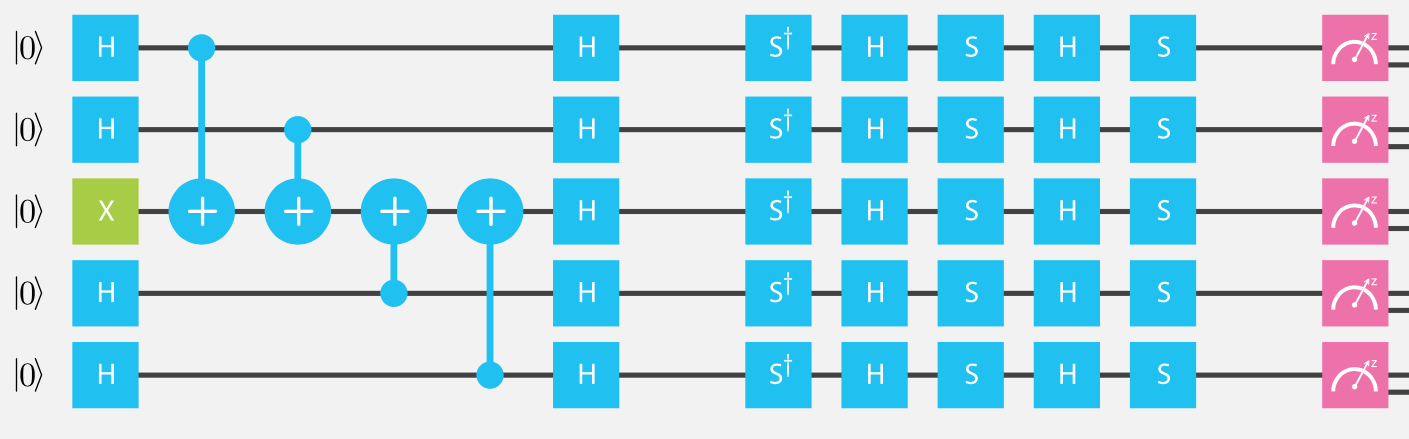}};

\def \Xs {3}; 
\def \Ys {-1.8}; 
\fill[white,draw=black] (\Xs+0.3, \Ys+2.3) rectangle (\Xs+1.1,\Ys+5.7);
\node at (0.9+\Xs,4+\Ys){   \includegraphics[height=3cm]{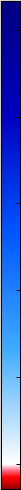}%
};
\node[align=right,anchor=east] at (0.8+\Xs,4+0.8+\Ys) {\footnotesize$2$};
\node[align=right,anchor=east] at (0.8+\Xs,4-0.3+\Ys) {\footnotesize$1$};
\node[align=right,anchor=east] at (0.8+\Xs,4-1.35+\Ys) {\footnotesize$0$};

 \def \centre {0.15}
 \def \A {1.35}
 \def \B {-0.2}
\draw[ultra thick,color = white] (2.2,-3) -- (\A,-3)-- (\A,\B); 
\draw[ultra thick,color = white,fill= white] (\A,\B) circle (1.2pt);
\draw[thick,color = black] (2.2,-3) -- (\A,-3)-- (\A,\B); 
\draw[thick,color = black,fill=black]  (\A,\B)circle (1pt); 

\draw[ultra thick,color = white] (-2,-3) -- (\centre,-3)-- (\centre,-2); 
\draw[ultra thick,color = white,fill= white] (\centre,-2) circle (1.2pt);
\draw[thick,color = black] (-2,-3) -- (\centre,-3)-- (\centre,-2); 
\draw[thick,color = black,fill=black]  (\centre,-2)circle (1pt); 

\draw[ultra thick,color = white,] (-2,3) -- (\centre,3)-- (\centre,1.3); 
\draw[ultra thick,color = white,fill= white] (\centre,1.3) circle (1.2pt);
\draw[thick,color = black] (-2,3) -- (\centre,3)-- (\centre,1.3); 
\draw[thick,color = black,fill=black] (\centre,1.3) circle (1pt); 

\draw[ultra thick,color = white,] (-3.6,-2) -- (-3.6,0.6)-- (-1.76,0.6); 
\draw[ultra thick,color = white,fill= white] (-1.76,0.6) circle (1.2pt);
\draw[thick,color = black] (-3.6,-2) -- (-3.6,0.6)-- (-1.76,0.6); 
\draw[thick,color = black,fill=black] (-1.76,0.6) circle (1pt); 


 \def \Lx {2.3}; 
 \def \Ly {0.5};
 \def \Tx {0.0}; 
 \def \Ty {0.24};
 \def \Bx {0.3}; 
 \def \Be {0.5}; 
 \def \By {0.17}; 

\definecolor{colourtwoSim}{rgb}{0.7893     0.90638           1}
\definecolor{colourtwoExp}{rgb}{0.80102     0.91159           1}

 \def \PAx {1.8};
 \def \PAy {-3};
\filldraw[fill=white, draw=black] (\PAx, {\PAy-\Ly}) rectangle ({\PAx+\Lx},{\PAy+\Ly});
\node[align=left,anchor=west] at  ({\PAx+\Tx}, {\PAy+\Ty})  {\footnotesize \sf Simulation};
\node[align=left,anchor=west] at  ({\PAx+\Tx}, {\PAy-\Ty})  {\footnotesize \sf Experiment};
\filldraw[fill= colourtwoSim, draw=black] ({(\PAx+\Lx)-\Be}, {\PAy+\Ty-\By}) rectangle ({(\PAx+\Lx)-\Be+\Bx},{\PAy+\Ty+\Bx-\By});
\filldraw[fill= colourtwoExp, draw=black] ({(\PAx+\Lx)-\Be}, {\PAy-\Ty+\By}) rectangle ({(\PAx+\Lx)-\Be+\Bx},{\PAy-\Ty-\Bx+\By});

\definecolor{colourfourSim}{rgb}{0     0.15106     0.86336}
\definecolor{colourfourExp}{rgb}{ 0     0.24714     0.93542}

 \def \PAx {-3};
 \def \PAy {-3};
\filldraw[fill=white, draw=black] (\PAx, {\PAy-\Ly}) rectangle ({\PAx+\Lx},{\PAy+\Ly});
\node[align=left,anchor=west] at  ({\PAx+\Tx}, {\PAy+\Ty})  {\footnotesize \sf Simulation};
\node[align=left,anchor=west] at  ({\PAx+\Tx}, {\PAy-\Ty})  {\footnotesize \sf Experiment};
\filldraw[fill= colourfourSim, draw=black] ({(\PAx+\Lx)-\Be}, {\PAy+\Ty-\By}) rectangle ({(\PAx+\Lx)-\Be+\Bx},{\PAy+\Ty+\Bx-\By});
\filldraw[fill= colourfourExp, draw=black] ({(\PAx+\Lx)-\Be}, {\PAy-\Ty+\By}) rectangle ({(\PAx+\Lx)-\Be+\Bx},{\PAy-\Ty-\Bx+\By});

\definecolor{colouroneSim}{rgb}{0      0.1787     0.88409}
\definecolor{colouroneExp}{rgb}{0     0.29489     0.97124}
 \def \PAx {-3.2};
 \def \PAy {3};
\filldraw[fill=white, draw=black] (\PAx, {\PAy-\Ly}) rectangle ({\PAx+\Lx},{\PAy+\Ly});
\node[align=left,anchor=west] at  ({\PAx+\Tx}, {\PAy+\Ty})  {\footnotesize \sf Simulation};
\node[align=left,anchor=west] at  ({\PAx+\Tx}, {\PAy-\Ty})  {\footnotesize \sf Experiment};
\filldraw[fill= colouroneSim, draw=black] ({(\PAx+\Lx)-\Be}, {\PAy+\Ty-\By}) rectangle ({(\PAx+\Lx)-\Be+\Bx},{\PAy+\Ty+\Bx-\By});
\filldraw[fill= colouroneExp, draw=black] ({(\PAx+\Lx)-\Be}, {\PAy-\Ty+\By}) rectangle ({(\PAx+\Lx)-\Be+\Bx},{\PAy-\Ty-\Bx+\By});

 \def \PAx {-4.2};
 \def \PAy {-1.6};
 \definecolor{colourthreeSim}{rgb}{1    0.069291    0.086275}
 \definecolor{colourthreeExp}{rgb}{1     0.86772      0.8902}

\filldraw[fill=white, draw=black] (\PAx, {\PAy-\Ly}) rectangle ({\PAx+\Lx},{\PAy+\Ly});
\node[align=left,anchor=west] at  ({\PAx+\Tx}, {\PAy+\Ty})  {\footnotesize \sf Simulation};
\node[align=left,anchor=west] at  ({\PAx+\Tx}, {\PAy-\Ty})  {\footnotesize \sf Experiment};
\filldraw[fill= colourthreeSim, draw=black] ({(\PAx+\Lx)-\Be}, {\PAy+\Ty-\By}) rectangle ({(\PAx+\Lx)-\Be+\Bx},{\PAy+\Ty+\Bx-\By});
\filldraw[fill= colourthreeExp, draw=black] ({(\PAx+\Lx)-\Be}, {\PAy-\Ty+\By}) rectangle ({(\PAx+\Lx)-\Be+\Bx},{\PAy-\Ty-\Bx+\By});

 \def \UB {-6.7};

\node[anchor=center] at (-2.25,\UB) {$\underbrace{\hspace{90pt}}_{\text{\footnotesize \sf Gates for \ket{\mathrm{GHZ}}}}$};
\node[anchor=center] at (1.6,\UB) {$\underbrace{\hspace{85pt}}_{\text{\ \footnotesize \sf Gates performing rotations}}$};
\node[anchor=center] at (3.95,\UB) {$\underbrace{\hspace{10pt}}_{\text{\ \footnotesize \sf Output}}$};

\end{tikzpicture}

\caption{\label{fig:spin_cat} Here we show the five-qubit GHZ spin Schr\"odinger cat state Wigner function $W_{\SU{2^5}}$ for the $\theta_1=\theta_2=\theta_3=\theta_4=\theta_5$ and $\varphi_1=\varphi_2=\varphi_3=\varphi_4=\varphi_5$ slice. 
  This can be considered a qubit-system analogue of~\Fig{fig:classic} and which was presented in~\cite{Deleglise:2008gt} to reconstruct non-classical cavity field states.  
  We note that in~\cite{Deleglise:2008gt} the interference terms that were observed correspond to quantum coherence in macroscopically distinct superpositions of states. 
  In this figure, the interference terms should be interpreted as a direct visualisation of the entanglement in the system. 
  Here we show the ideal function, and as insets, show both simulated and experimental results from IBM's \emph{Quantum Experience} project. 
  In this figure we also show an example circuit used to generate simulated and experimental data. 
  As with the circuits used to create the Bell states presented in~\Fig{fig:IBM2qubit}, these gate operations ideally would be replaced by optimized, single-rotation, operations that have very recently been made available by IBM. 
  We note that the two, non-polar, points can be obtained in a variety of ways. 
  Specifically they could be found by using just $\theta$ rotations, or through a combination of $\theta$ and $\varphi$ rotations. 
  We have verified that the results that we obtained from the IBM \emph{Quantum Experience} project are independent of the combination of rotations used.
    }
\end{figure}
Bell states are interesting both as an example of maximally entangled states and for their usefulness in quantum information processing.
Fortunately, for systems comprising more spins, we can extend this class of states to those that have a direct analogy with optical Schr\"odinger cat states as considered in~\cite{Deleglise:2008gt} and others. 
Such states are termed ``spin-cat states'' of which the GHZ state~\cite{GHZ} is an excellent example. 
In previous theoretical work, spin Wigner-like functions have been proposed as a mechanism for visualizing such cat states~\cite{PhysRevA.49.4101,PhysRevA.85.022113,PhysRevA.87.052323}. 
In analogy with measuring Wigner functions of non-classical cavity field states~\cite{Deleglise:2008gt}, using our method we now construct the $W_{\SU{2^{[5]}}}$ Wigner function for a spin-cat of the form $\ket{\mathrm{GHZ}_5}=(\ket{0}_1\ket{0}_2 \ket{0}_3\ket{0}_4\ket{0}_5+\ket{1}_1\ket{1}_2 \ket{1}_3\ket{1}_4\ket{1}_5)/\sqrt{2}$.
In~\Fig{fig:spin_cat} we show the $\theta_1=\theta_2=\theta_3=\theta_4=\theta_5$ and $\varphi_1=\varphi_2=\varphi_3=\varphi_4=\varphi_5$ slice of the $W_{\SU{2^{[5]}}}$ Wigner function for $\ket{\mathrm{GHZ}_5}$ which is the higher dimensional analogue of~\Fig{fig:example}(a,b). 
We show both theoretical predictions and, due to limited computational resource, as insets, simulation and experimental data obtained from the IBM machine. 
Once more the calibration data pertaining to the experiments is provided in Table~\ref{table}.
We note that the $\theta_1=\theta_2=\theta_3=\theta_4=\theta_5$ and $\varphi_1=\varphi_2=\varphi_3=\varphi_4=\varphi_5$ slice does not contain all the information needed to reconstruct the state; for full reconstruction we would need to measure and visualise all $\{\theta_i, \theta_j\} \, i \neq j$ sets of angles for various values of $\varphi_i$. 
For the top and bottom point the theoretical value is $2.7$ while the simulated values are $1.64$ and $1.70$, and experimental values $1.16$ and $1.22$, respectively. 
Here simulation and experiment are in good agreement. 
The difference from the theoretical values for all four points indicates that there is some decoherence and/or gate and measurement errors in the system, mostly accounted for in IBM's simulation, meaning that the observed state is not in an ideal GHZ state. 
 
\section{A Wigner function for tensor products of spins \label{TPsec}}
The Stratonovich-Weyl conditions do not uniquely specify the extended parity operator $\Pi$ and hence the Wigner function is also not uniquely defined. 
Because of this, it is natural to ask what difference choosing alternative Wigner functions will make. 
As our current focus is on experimental reconstruction of the quantum state in phase space, we believe that it is instructive to explore at least one alternative whose direct measurement may be more readily available to those working in quantum information.
In the previous case, the definition of extended parity was motivated by the idea of a global $\pi$ rotation on the hypersphere of the underlying \SU{2^{[N]}} coherent state representation.
In this case the notion of extended parity is motivated on an individual qubit level; a global $\pi$ rotation on each qubit's Bloch sphere.
This leads to a extended parity operator that is nothing more than the tensor product of the parities of individual qubits:
\be
\Pi_{\bigotimes ^N \SU{2}}=\bigotimes_{i=1}^N  \Pi_{\SU{2^{[1]}}}^{(i)}=\bigotimes_{i=1}^N\Half(\Bid+\sqrt{3}\sigma_{z_i}),
\ee 
which for one qubit is equal to~\Eq{pione} but for two qubits takes the explicit form
\ba 
\Pi_{\bigotimes ^2 \SU{2}} &=&\frac{1}{2}
\begin{pmatrix}
2+\sqrt{3} & 0 & 0 & 0 \\ 
0&-1 & 0 & 0 \\ 
0&0&-1 & 0  \\ 
 0 & 0 & 0 &2-\sqrt{3}\\ 
\end{pmatrix} \\
&=& \frac{1}{4}[\Bid \otimes \Bid + \sqrt{3} \, \Bid \otimes \sigma_z + \sqrt{3} \, \sigma_z \otimes \Bid + 3 \, \sigma_z \otimes \sigma_z] \nonumber 
\ea
in the computational basis. 
When compared with~\Eq{parity} we see that this version of extended parity no longer treats one-qubit and two-qubit contributions on an equal footing.
The definition of the Wigner function continues in the same way as before and, in terms of the rotated density matrix $\tilde{\rho}=\mathbb{U}_N^\dag \rho \mathbb{U}_N$ takes the form
\ba
W_{\bigotimes ^N \SU{2}}(\Param)&=&\Trace{\tilde\DO(\Param) \Pi_{\bigotimes ^N \SU{2}}}
\nonumber \\ \label{wprime}
&=&\sum_n{\tilde\DO_{nn}(\Param)}\left(\Pi_{\bigotimes ^N \SU{2}}\right)_{nn}.
\ea
Returning to~\Fig{fig:example}(e,f) we show example slices of $W_{\bigotimes ^N \SU{2}}^{\theta_i=0}(\theta_1,\theta_2)=W_{\bigotimes ^N \SU{2}}(\theta_1,\varphi_1=0,\theta_2,\varphi_2=0)$  that demonstrates this alternative Wigner function is qualitatively very similar to the equivalent slices of $W_{\SU{2^{[N]}}}(\Omega)$ shown in~\Fig{fig:example}(c,d). 
 \begin{figure}[!t]
\begin{tikzpicture}[scale=1.0]
\def \fy {-7.3}
\node[inner sep=0pt] at (0,0)
    {\includegraphics[width=\linewidth]{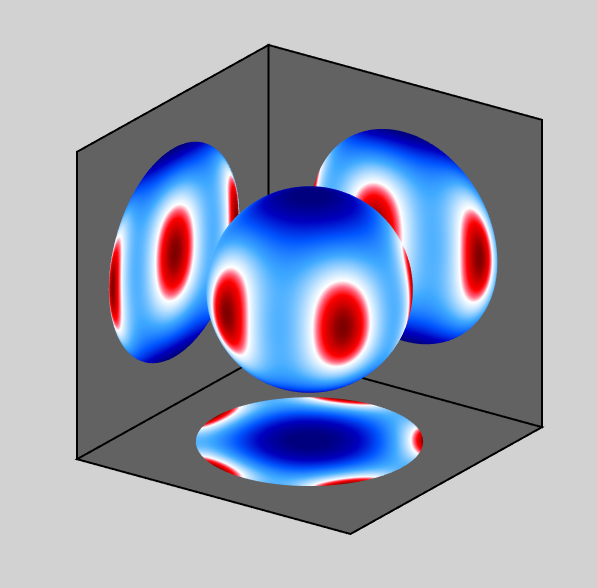}};
\node[inner sep=0pt] at (0, \fy)
    {\includegraphics[width=\linewidth]{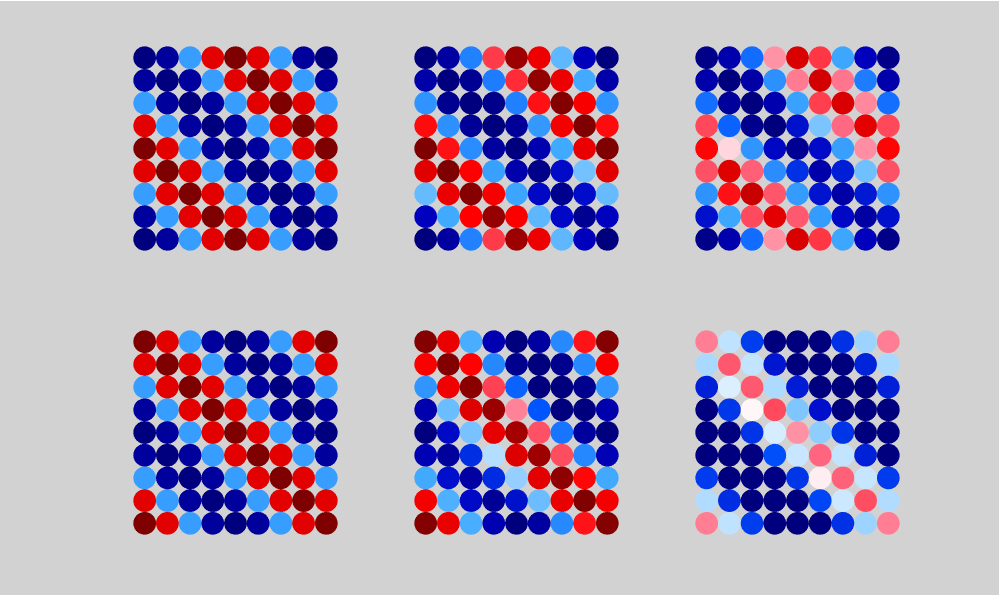}};
\def \Xs {3}; 
\def \Ys {-1.8}; 
\fill[white,draw=black] (\Xs+0.3, \Ys+2.3) rectangle (\Xs+1.1,\Ys+5.7);
\node at (0.9+\Xs,4+\Ys){   \includegraphics[height=3cm]{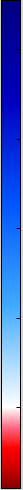}%
};
\node[align=right,anchor=east] at (0.8+\Xs,4+0.1+\Ys) {\footnotesize$1$};
\node[align=right,anchor=east] at (0.8+\Xs,4+1.2+\Ys) {\footnotesize$2$};
\node[align=right,anchor=east] at (0.8+\Xs,4-1+\Ys) {\footnotesize$0$};

 \def \centre {0.15}
 \def \A {1.35}
 \def \B {-0.2}
\draw[ultra thick,color = white] (2.2,-3) -- (\A,-3)-- (\A,\B); 
\draw[ultra thick,color = white,fill= white] (\A,\B) circle (1.2pt);
\draw[thick,color = black] (2.2,-3) -- (\A,-3)-- (\A,\B); 
\draw[thick,color = black,fill=black]  (\A,\B)circle (1pt); 

\draw[ultra thick,color = white] (-2,-3) -- (\centre,-3)-- (\centre,-2); 
\draw[ultra thick,color = white,fill= white] (\centre,-2) circle (1.2pt);
\draw[thick,color = black] (-2,-3) -- (\centre,-3)-- (\centre,-2); 
\draw[thick,color = black,fill=black]  (\centre,-2)circle (1pt); 

\draw[ultra thick,color = white,] (-2,3) -- (\centre,3)-- (\centre,1.3); 
\draw[ultra thick,color = white,fill= white] (\centre,1.3) circle (1.2pt);
\draw[thick,color = black] (-2,3) -- (\centre,3)-- (\centre,1.3); 
\draw[thick,color = black,fill=black] (\centre,1.3) circle (1pt); 

\draw[ultra thick,color = white,] (-3.6,-2) -- (-3.6,0.6)-- (-1.76,0.6); 
\draw[ultra thick,color = white,fill= white] (-1.76,0.6) circle (1.2pt);
\draw[thick,color = black] (-3.6,-2) -- (-3.6,0.6)-- (-1.76,0.6); 
\draw[thick,color = black,fill=black] (-1.76,0.6) circle (1pt); 


 \def \Lx {2.3}; 
 \def \Ly {0.5};
 \def \Tx {0.0}; 
 \def \Ty {0.24};
 \def \Bx {0.3}; 
 \def \Be {0.5}; 
 \def \By {0.17}; 

\definecolor{colourtwoSim}{rgb}{0.20303     0.60391           1}
\definecolor{colourtwoExp}{rgb}{0.30137      0.6899           1}

 \def \PAx {1.8};
 \def \PAy {-3};
\filldraw[fill=white, draw=black] (\PAx, {\PAy-\Ly}) rectangle ({\PAx+\Lx},{\PAy+\Ly});
\node[align=left,anchor=west] at  ({\PAx+\Tx}, {\PAy+\Ty})  {\footnotesize \sf Simulation};
\node[align=left,anchor=west] at  ({\PAx+\Tx}, {\PAy-\Ty})  {\footnotesize \sf Experiment};
\filldraw[fill= colourtwoSim, draw=black] ({(\PAx+\Lx)-\Be}, {\PAy+\Ty-\By}) rectangle ({(\PAx+\Lx)-\Be+\Bx},{\PAy+\Ty+\Bx-\By});
\filldraw[fill= colourtwoExp, draw=black] ({(\PAx+\Lx)-\Be}, {\PAy-\Ty+\By}) rectangle ({(\PAx+\Lx)-\Be+\Bx},{\PAy-\Ty-\Bx+\By});

\definecolor{colourfourSim}{rgb}{0           0         0.5}
\definecolor{colourfourExp}{rgb}{ 0           0         0.5}

 \def \PAx {-3};
 \def \PAy {-3};
\filldraw[fill=white, draw=black] (\PAx, {\PAy-\Ly}) rectangle ({\PAx+\Lx},{\PAy+\Ly});
\node[align=left,anchor=west] at  ({\PAx+\Tx}, {\PAy+\Ty})  {\footnotesize \sf Simulation};
\node[align=left,anchor=west] at  ({\PAx+\Tx}, {\PAy-\Ty})  {\footnotesize \sf Experiment};
\filldraw[fill= colourfourSim, draw=black] ({(\PAx+\Lx)-\Be}, {\PAy+\Ty-\By}) rectangle ({(\PAx+\Lx)-\Be+\Bx},{\PAy+\Ty+\Bx-\By});
\filldraw[fill= colourfourExp, draw=black] ({(\PAx+\Lx)-\Be}, {\PAy-\Ty+\By}) rectangle ({(\PAx+\Lx)-\Be+\Bx},{\PAy-\Ty-\Bx+\By});

\definecolor{colouroneSim}{rgb}{0           0         0.5}
\definecolor{colouroneExp}{rgb}{0           0     0.53959}
 \def \PAx {-3.2};
 \def \PAy {3};
\filldraw[fill=white, draw=black] (\PAx, {\PAy-\Ly}) rectangle ({\PAx+\Lx},{\PAy+\Ly});
\node[align=left,anchor=west] at  ({\PAx+\Tx}, {\PAy+\Ty})  {\footnotesize \sf Simulation};
\node[align=left,anchor=west] at  ({\PAx+\Tx}, {\PAy-\Ty})  {\footnotesize \sf Experiment};
\filldraw[fill= colouroneSim, draw=black] ({(\PAx+\Lx)-\Be}, {\PAy+\Ty-\By}) rectangle ({(\PAx+\Lx)-\Be+\Bx},{\PAy+\Ty+\Bx-\By});
\filldraw[fill= colouroneExp, draw=black] ({(\PAx+\Lx)-\Be}, {\PAy-\Ty+\By}) rectangle ({(\PAx+\Lx)-\Be+\Bx},{\PAy-\Ty-\Bx+\By});

 \def \PAx {-4.2};
 \def \PAy {-1.6};
 \definecolor{colourthreeSim}{rgb}{0.8           0           0}
 \definecolor{colourthreeExp}{rgb}{1     0.30551     0.38039}

\filldraw[fill=white, draw=black] (\PAx, {\PAy-\Ly}) rectangle ({\PAx+\Lx},{\PAy+\Ly});
\node[align=left,anchor=west] at  ({\PAx+\Tx}, {\PAy+\Ty})  {\footnotesize \sf Simulation};
\node[align=left,anchor=west] at  ({\PAx+\Tx}, {\PAy-\Ty})  {\footnotesize \sf Experiment};
\filldraw[fill= colourthreeSim, draw=black] ({(\PAx+\Lx)-\Be}, {\PAy+\Ty-\By}) rectangle ({(\PAx+\Lx)-\Be+\Bx},{\PAy+\Ty+\Bx-\By});
\filldraw[fill= colourthreeExp, draw=black] ({(\PAx+\Lx)-\Be}, {\PAy-\Ty+\By}) rectangle ({(\PAx+\Lx)-\Be+\Bx},{\PAy-\Ty-\Bx+\By});

 \def \UB {-6.7};

\def \dX {0.1}
\node[anchor=west] at (0,0.2+\fy) {$\theta_1$};
\node[anchor=west] at (-2.5,0.2+\fy) {$\theta_1$};
\node[anchor=west] at (2.5,0.2+\fy) {$\theta_1$};
\node[anchor=west] at (0,-2.2+\fy) {$\theta_1$};
\node[anchor=west] at (-2.5,-2.2+\fy) {$\theta_1$};
\node[anchor=west] at (2.5,-2.2+\fy) {$\theta_1$};
\node[anchor=west] at (-3.6-\dX,-1.1+\fy) {$\theta_2$};
\node[anchor=west] at (-1.15-\dX,-1.1+\fy) {$\theta_2$};
\node[anchor=west] at (1.3-\dX,-1.1+\fy) {$\theta_2$};
\node[anchor=west] at (-3.6-\dX,1.3+\fy) {$\theta_2$};
\node[anchor=west] at (-1.15-\dX,1.3+\fy) {$\theta_2$};
\node[anchor=west] at (1.3-\dX,1.3+\fy) {$\theta_2$};

\node[anchor=west,rectangle,draw, fill=white] at (-2.9,2.6+\fy) {\sf Theory};
\node[anchor=west,rectangle,draw, fill=white] at (-0.66,2.6+\fy) {\sf Simulated};
\node[anchor=west,rectangle,draw, fill=white] at (1.6,2.6+\fy) {\sf Experimental};

\def \Xs {3.3}; 
\def \Ys {-4.7}; 
\node at (0.9+\Xs,4+\Ys+\fy){   \includegraphics[height=3cm]{colourbar}%
};
\node[align=right,anchor=east] at (0.85+\Xs,2.5+\Ys+\fy) {\footnotesize$-0.5$};
\node[align=right,anchor=east] at (0.85+\Xs,4.5+\Ys+\fy) {\footnotesize$0.5$};
\node[align=right,anchor=east] at (0.85+\Xs,5.45+\Ys+\fy) {\footnotesize$1$};
\node[align=right,anchor=east] at (0.85+\Xs,3.5+\Ys+\fy) {\footnotesize$0$};

\node[anchor=west,rectangle,draw, fill=white] at (-4.3,1.8+\fy) {$\ket{\Phi_-}$};
\node[anchor=west,rectangle,draw, fill=white] at (-4.3,-0.5+\fy) {$\ket{\Psi_+}$};

\end{tikzpicture}

\caption{\label{fig:prime} 
Here we reproduce Figs.~3 and~4 using the same data but now employing the Wigner function defined using the alternative extended parity operators as given in~\Eq{wprime}. 
In the top figure, for comparison with Fig.~4, we show the five-qubit GHZ spin Schr\"odinger cat state Wigner function $W_{\bigotimes^5 \SU{2}}$ for the $\theta_1=\theta_2=\theta_3=\theta_4=\theta_5$ and $\varphi_1=\varphi_2=\varphi_3=\varphi_4=\varphi_5$ slice. 
Again we show the ideal function, and as insets, show both simulated and experimental results from IBM's \emph{Quantum Experience} project. 
On the bottom figure, for comparison with Fig.~3, we provide plots of $W_{\bigotimes^2 \SU{2}}$ for the two Bell states $\ket{\Phi_+}$ and $\ket{\Psi_-}$. 
We plot $\theta_1$ versus $\theta_2$ for the $\varphi_1=\varphi_2=0$ slice of the Wigner function for two qubits. 
Once more, we have included for comparison ideal theoretical values, numerical results using IBM's built in simulator, and real experimental data from IBM's quantum processor. 
Again we see good agreement between theory, simulation and experiment and note that using a different extended parity operator provides an alternative path to direct measurement of phase space.}
\end{figure}
 
In~\Fig{fig:prime} (top) we show results for comparison with~\Fig{fig:spin_cat} and (bottom) (and, by analogy, with non-classical cavity field states~\cite{Deleglise:2008gt}) with~\Fig{fig:IBM2qubit} which demonstrates that $W_{\bigotimes ^N \SU{2}}$ is a Wigner function with qualitatively very similar features to $W_{\SU{2^{[N]}}}$ that will be compared in the next section. 
For the top and bottom point the theoretical value is $2.375$. 
The simulated values are $1.13$ and $1.11$, and the experimental values  are $0.8876$ and $0.9006$, respectively. 

\section{Efficient state estimation, characterisation, and entanglement validation}
As they are informationally complete, our Wigner functions for spin can be considered mathematically equivalent to the density matrix/state space formulation.  They also exhibit unique, and intuitively natural, characteristic features. If, for example, we look at~\Fig{fig:prime} for the GHZ state (which is a superposition of spin coherent states) it is clear that there are regions of strong oscillations in the equal angle slice; these are reminiscent of the interference terms between two harmonic oscillator coherent states shown in~\Fig{fig:classic}. It is natural to ask if measurement of such characteristic features can be used to verify non-classical properties of the state such as quantum coherence or entanglement. In other words, can we extract information in a similar way as for Wigner functions of continuous systems where negativity is a signature of non-classical correlations? In finite dimensional systems things are a little more complicated as negativity of the Wigner function has some subtle complexities which we will expand on later in this manuscript and in full detail in a later work.
Moreover, the exact form of a state's spin Wigner function is fixed by the chosen extended parity operator that is used.
As such, it may be that different extended parity operations may be more or less useful in revealing particular characteristic features of the quantum state. In order to focus the discussion in this section we fix our choice of parity and Wigner function to $\Pi_{\bigotimes ^N \SU{2}}$ and $W_{\bigotimes ^N \SU{2}}$.
We discuss with reference to this specific Wigner function possibilities for efficient state characterisation/categorisation (e.g. by identifying features peculiar to GHZ states). 
We show that if one has sufficient prior information about the expected state of the system (such as, that it comprises a superposition of antipodal spin coherent states) it may be possible to validate entanglement with only a couple of measurements.

To begin we consider the $N$-qubit state
\be
\label{rhoGHZ}
\rho(\gamma) = \gamma \rho_\text{GHZ} + ( 1 - \gamma ) \rho_{m}
\ee
where $\gamma\in[0,1]$. Here $\rho(\gamma)$ interpolates between the  density operators $\rho_\mathrm{GHZ}$ for the GHZ state (the coherent superposition of \ket{11111} and \ket{00000} with $\gamma=1$) and $\rho_{m}$ for the statistical mixture of \ket{11111} and \ket{00000} (with $\gamma=0$).
The Wigner function of this state is 
\ba 
\label{analGHZ}
W_{\bigotimes ^N \SU{2}}^{(\gamma)}(\Param)&&=\frac{1}{2^{N+1}}\prod_{i=1}^{N}(1+\sqrt{3}\cos 2\theta_i)  \\ 
&&+ \frac{1}{2^{N+1}}\prod_{i=1}^{N}(1-\sqrt{3}\cos 2\theta_i)   \nonumber \\
&&+ \frac{\gamma}{2^{N}}\prod_{i=1}^{N}(\sqrt{3}\sin 2\theta_i)\cos\left(2\sum_i^N{\varphi_i }\right). \nonumber
\ea
When $\gamma = 1$ we can see that the $N$-qubit GHZ state is made up of three terms: the first two correspond to the first and last diagonal elements of the density matrix in~\Eq{rhoGHZ} and the third (interference) term to the maximally off-diagonal elements.
\Fig{fig:mixedVentangled} compares the equal-angle Wigner functions $(\theta=\theta_{1}=\cdots=\theta_{N}, \varphi=\varphi_{1}=\cdots=\varphi_{N})$ of (a) the GHZ state $\gamma=1$ and (b) the separable mixed state $\gamma=0$.
As can be seen, the maxima at the top and bottom of the sphere are the same in both states, although the equatorial oscillations are absent in the separable state.

From this simple example, it is clear that the oscillations around the equator, where all $\theta_i = \pi/4$, arise entirely from the $\cos\left(2\sum_{i}^{N}\varphi_i\right)$ term.  
These oscillations, which are of maximum possible frequency for a Wigner function with this number of qubits, are characteristic of GHZ type superposition (compare the iconic Wigner function~\Fig{fig:classic}) and are analogous to the super-resolution oscillations observed in $N00N$ states \cite{Boto}. We note that any antipodal superposition of spin coherent states will be look like a rotated version of  ~\Fig{fig:mixedVentangled}(a) with interference terms along the geodesic bisecting them.  It is natural to ask if such oscillations can be used to certify GHZ-type entanglement. We note that negativity in the Wigner function alone is insufficient to be a signal of entanglement. To illustrate this we show in~\Fig{fig:mixedVentangled}(c) the equal angle slice Wigner function for the state \ket{10000} and note that despite being separable it has significant negativity in this equal-angle slice. Indeed the equal-angle slice of $W_{\bigotimes^5 \SU{2}}$ function for the statistical mixture of $\ket{10000}$, $\ket{01000}$, etc is identical to \Fig{fig:mixedVentangled}(c). In order to establish if there is a potential to use the characteristic features of the GHZ Wigner function equal-angle slice for certification we can ask what is the nearest separable state in terms of its phase space characteristics. We believe the closest in form is the `clock' state which we define by 
\bel{clockstate}\ket{\psi_\text{clock}}=\frac{1}{2^{N/2}}\bigotimes_{k=1}^{N}\left[\ket0+\exp\left({2\ui \pi k\over N}\right)\ket1 \right],\ee 
whose Wigner function is
\be
\label{clock}
W_{\bigotimes ^N \SU{2}}^\text{clock}(\Param)
=\frac{1}{2^N}\prod_{k=1}^N 1+\sqrt{3}\sin 2\theta_k \cos\left(
2\varphi_k+\frac{2\pi k}{N}\right)\ee
We show the equal angle slice of this function in \Fig{fig:mixedVentangled}(d). We note that there is a similar oscillatory character to that seen in the GHZ state but that it is exponentially smaller in amplitude. For this reason we show this function again in \Fig{fig:mixedVentangled}(e) but on a different scale. It is straightforward to show\footnote{The maximum-frequency equatorial oscillations of the Wigner function are determined by the top-right and bottom-left elements of the density matrix.  The maximum amplitude of these for any product state $\bigotimes_{k=1}^N (a_k\ket0 + b_k\ket1)$ occurs when $|a_k|=|b_k|=1/\sqrt2$ and has magnitude $2^{-N}$, compared with $2^{-1}$ for the GHZ state.} therefore that oscillations of this wavelength that exceed those of the clock state Wigner oscillations is a signature of a GHZ type of entanglement - something that in principle can be established with only two measurements.
\begin{figure*}[!t]
\resizebox{\linewidth}{!}{
\begin{tikzpicture}[scale=1]
\node[anchor=south] at (0,0) {
  \includegraphics[width=\linewidth,trim={100 50 30 30},clip = true]{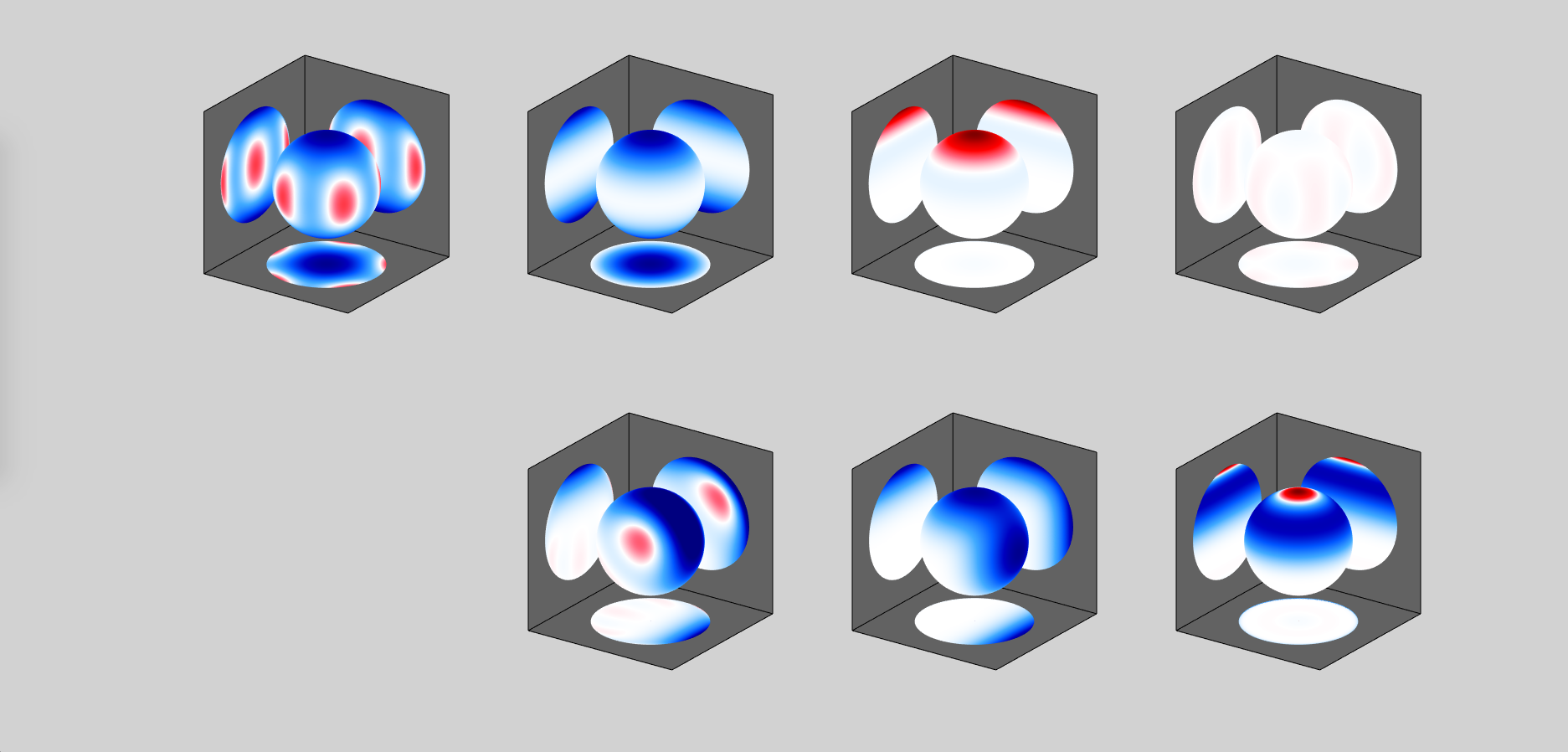}
};
\node[anchor=south] at (-5.9,0.3) {
  \includegraphics[width=0.185\linewidth,trim={110 30 90 20},clip = true]{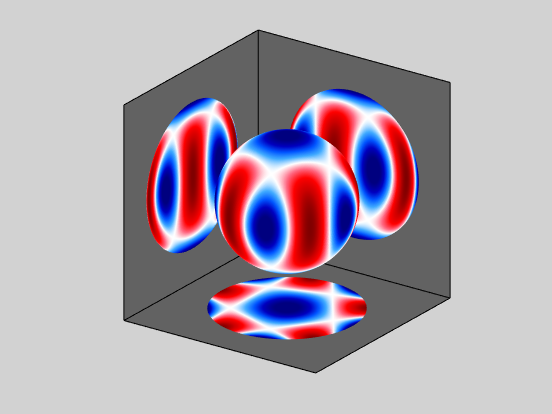}
};

\node at (-7.85,2.1){   \includegraphics[height=3cm]{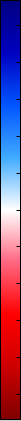}%
};
\node[align=right,anchor=east] at (-7.9,0.8) {\footnotesize$-0.025$};
\node[align=right,anchor=east] at (-7.9,2.1) {\footnotesize$0$};
\node[align=right,anchor=east] at (-7.9,3.4) {\footnotesize$0.025$};
\draw[line width=1mm, color=white] (-9,4.4) -- (-3.9, 4.4) -- (-3.9, 0.1);
\def \Xxs {7.5};
\def \Yys {0.5};

\node at (0.9+\Xxs,4+\Yys){   \includegraphics[height=3cm]{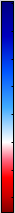}%
};
\node[align=right,anchor=east] at (0.8+\Xxs,4-1.3+\Yys) {\footnotesize$-1$};
\node[align=right,anchor=east] at (0.8+\Xxs,4-0.5+\Yys) {\footnotesize$0$};
\node[align=right,anchor=east] at (0.8+\Xxs,4+0.25+\Yys) {\footnotesize$1$};
\node[align=right,anchor=east] at (0.8+\Xxs,4+1.05+\Yys) {\footnotesize$2$};
\node[align=right,anchor=east] at (-6.9,8.1) {(a)};
\node[align=right,anchor=east] at (-2.9,8.1) {(b)};
\node[align=right,anchor=east] at (1.1,8.1) {(c)};
\node[align=right,anchor=east] at ( 5.1,8.1) {(d)};
\node[align=right,anchor=east] at (-6.9,3.7) {(e)};
\node[align=right,anchor=east] at (-2.9,3.7) {(f)};
\node[align=right,anchor=east] at (1.1,3.7) {(g)};
\node[align=right,anchor=east] at (5.1,3.7) {(h)};
\end{tikzpicture}}
    \caption{\label{fig:mixedVentangled} Here we show the equal angle slice $W_{\bigotimes ^N \SU{2}}$ Wigner function for various 5-qubit states. (a) shows the same GHZ state as seen in \Fig{fig:prime} with (b) showing the mixed state counterpart of this GHZ state given by $\left((\ket{0}\bra{0})^{\otimes N} + (\ket{1}\bra{1})^{\otimes N} \right)/{2}$ Only the pure state displays the interference pattern given by the off diagonal terms in the density matrix when the state is entangled. In (c) we see the state $\ket{10000}$. In figures (d-e) we see the clock state, (d) is shown with the same colour map as the other plots, whereas (e) shows the state with a modified colour map to show the structure of the slice that is not evident with the colour maps used throughout the rest of this figure. (f-g) Defining $\ket{\rightarrow}=(\ket{0}+\ket{1})/\sqrt{2}$ we show the entangled superposition of spin coherent states $\ket{0}^{\otimes N} + \ket{\rightarrow}^{\otimes N}$ and its mixed state counterpart, the equally weighted statistical mixture of $(\ket{0}\bra{0})^{\otimes N}$ and $(\ket{\rightarrow}\bra{\rightarrow})^{\otimes N}$.  Again note that only the pure state has negative interference terms in this slice with two of particularly large amplitude. Finally (h) shows the equal angle slice Wigner function for the five-qubit $W$-state showing that other entangled states have patterns that could also act as state identification signatures.
}
\end{figure*}
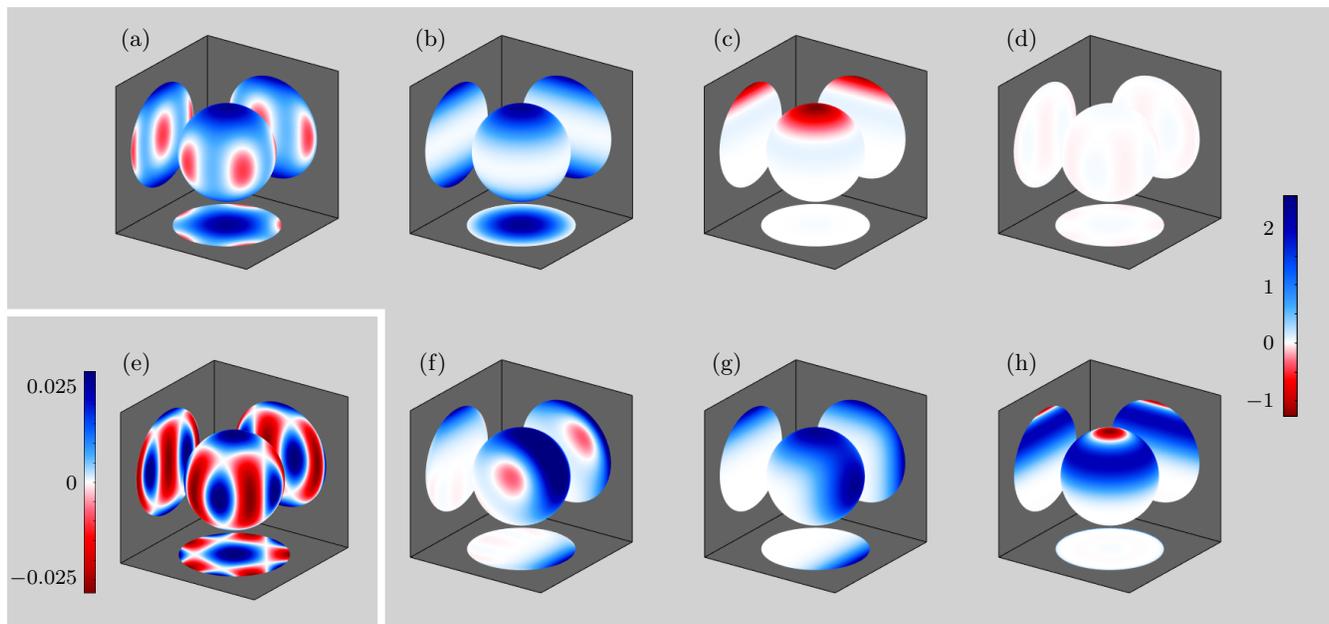
\begin{figure}[!t]
   \includegraphics[trim={4.0cm 13.1cm 3.4cm 4.8cm},clip,width=\linewidth]{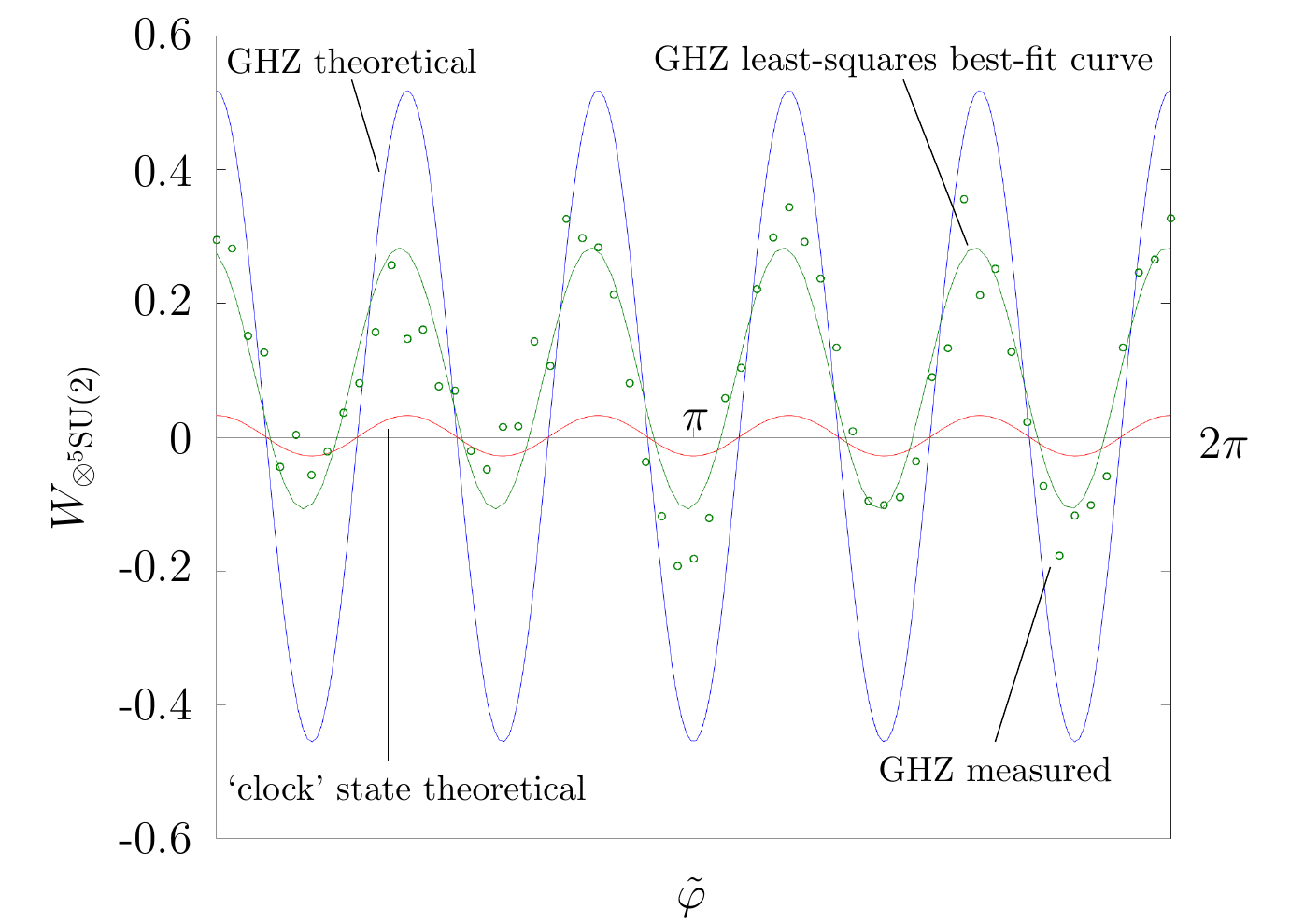}
    \caption{\label{fig:equatorial} 
    Points around the equator of the 5 qubit GHZ state Wigner function from~\Fig{fig:prime} with $\theta_{1}=\theta_{2}=\cdots=\theta_{N}=\frac\pi4, \varphi_{1}=\varphi_{2}=\cdots=\varphi_{N}=\varphi$ where the blue curve is the theoretically calculated values for an ideal GHZ state found from our model. 
The green dots are the measured experimental results and the green curve is a least-square best-fit sinusoid to the experimental results. 
In red we see the oscillations given around the equator for the separable `clock' state of~\Eq{clockstate};  the amplitude is significantly lower than for the ``GHZ measured'' state  demonstrating the latter (whose oscillations are not as strong as the theoretical maximum) is indeed entangled. Here $\tilde{\varphi}= 2\varphi$ to correspond to how IBM define the rotations on their machine. 
    }
\end{figure}
Due to an update on the IBM computer during the closing stages of our work, we were able to observe these oscillations directly as seen in~\Fig{fig:equatorial}.
This was due to the implementation of three new gates that can perform arbitrary rotations on individual qubits: $u_1(\lambda)$, $u_2(\tilde{\varphi},\lambda)$, and $u_3(\tilde{\theta},\tilde{\varphi},\lambda)$, with $u_3(\tilde{\theta},\tilde{\varphi},\lambda) = e^{-\ui \tilde{\varphi} \sigma_z/2}e^{-\ui \tilde{\theta} \sigma_y/2}e^{-\ui\lambda \sigma_z/2}$.
These three gates give us the freedom to specify any rotation or any point in phase space in \SU{2}, allowing us to sweep around the equator, experimentally verifying the presence of the interference-based oscillation for a 5-qubit GHZ state. 
In~\Fig{fig:equatorial} these measurement results are compared to ideal theoretical values.
The experimentally measured amplitudes are somewhat reduced, as well as having an offset phase.
This indicates that the computer is not producing a perfect GHZ state but that the state is verified to be both entangled and in reasonably consistent agreement with a perfect GHZ state. 
However, since there is a linear mapping between the density matrix and the Wigner function, a similar Wigner function implies the state is at least similar, making feature-based characterisation robust against small errors in state preparation and limited decoherence, likely candidates for the difference between the theoretical and experimental curves.

The advantage of our approach is in the potential to reduce the number of measurements required to develop confidence certifying more general states.
For example, begin by defining $\ket{\rightarrow}=(\ket{0}+\ket{1})/\sqrt{2}$.
We then generate the normalised equal superposition of $\ket{0}^{\otimes N}$ and $\ket{\rightarrow}^{\otimes N}$.
In~\Fig{fig:mixedVentangled} we show $W_{\bigotimes ^N \SU{2}}$ for this state (f) and the corresponding equally weighted mixture (g) of $\ket{0}^{\otimes N}$ and $\ket{\rightarrow}^{\otimes N}$.
Both density matrices have no non-zero elements in the computational basis (making conventional tomography challenging), but the superposition can be certified to be within an acceptable confidence interval through a few measurements of the characteristic features of its corresponding Wigner function.
As with our analysis of the GHZ state the presence of this structure may be used to give confidence in the existence of entanglement in the system and categorisation as a superposition of spin coherent states. As another example, we show in~\Fig{fig:mixedVentangled}(h) the equal angle Wigner function for the $W$-state of five qubits. 
Again we see that $W$-states have a distinctive shape (as $W$-states can be thought of as ``eigenstates of a total `$z$' angular momentum operator'' this form is intuitively sensible). 
Once more, it would not take more than a few measurements to gain significant confidence that a system was or was not in a $W$ state. 

In terms of the more general problem of rapid quantum state estimation spin-Wigner functions may be of particular use when some properties of the state are known in advance. 
We have already noted that only a few measurements are needed to verify certain characteristic features of the Wigner function are present. 
As it is possible to build these Wigner functions from  expansions using bases other than the computational basis such as from stabiliser states, full quantum-state reconstruction can be viewed as establishing the coefficients of such expansions. 
Understanding the structure of these expansions together with foreknowledge of the set of potential states a system may achieve can lead to efficient state estimation protocols. 
This is because one can select measurements that rapidly exclude very many of the components of the expansion and confirm the presence of the dominant terms. 
In this way phase space methods provide an alternative path to efficient state estimation from those known in other areas of quantum state tomography~\cite{Plenio2010,1612.08000,1367-2630-15-12-125004,PhysRevLett.111.020401,PhysRevLett.106.020401}. 
A detailed study of efficient quantum state reconstruction in phase space will be the subject of a future work.

\section{On the differences between Wigner functions}
Each of the two cases we have considered here have their own strengths which will be expanded on in a later publication. 
However, we are including a brief discussion to highlight that there is some freedom in choosing extended parity operators in tensor product spaces. 
This should be of utility as it increases the number of available options in designing experiments for the direct measurement of a Wigner function. 

The full-group Wigner function $W_{\SU{2^{[N]}}}$ and the tensor-product Wigner function $W_{\bigotimes^N \SU{2}}$ are related to the density matrix by different, but still invertible, linear maps, and therefore both contain full information about the quantum state.   
The tensor-product form has the additional property of respecting the marginals in each subspace. 
We can see this is indeed the case by noting that the two qubit kernel separates
\begin{equation}
\Delta_{\bigotimes^2 \SU{2}} = \Delta_{\SU{2}}(\Omega_{A})\otimes \Delta_{\SU{2}}(\Omega_{B})
\end{equation}
leading to the result
\ba
\int W_{\bigotimes^2 \SU{2}}(\Omega_{A},\Omega_{B}) \ud\Omega_{B} &=& \Trace{\rho_{A}  \Delta_{\SU{2}}(\Omega_{A})} \nonumber \\ 
&=& W_{\SU{2}}(\Omega_{A})
\ea
where $\rho_{A}$ is the reduced density matrix of subsystem $A$. Note that extension to arbitrary number of qubits is a trivial extension  of this argument. 

As an example, consider the Bell state $|\Psi_{+}\rangle$ shown in \Fig{fig:example}(b).  
Here our two Wigner function cases have the same structure, with the tensor-product form having a larger amplitude of modulation:
\begin{eqnarray}
W_{\SU{2^{[2]}}}  & = & \frac14 (1+\sqrt 5 (x_{A}x_{B}+y_{A}y_{B}-z_{A}z_{B})) \\
W_{\bigotimes^2 \SU{2}}  & = & \frac14 \left(1+3(x_{A}x_{B}+y_{A}y_{B}-z_{A}z_{B})\right), 
\end{eqnarray}
where $(x_{i},y_{i},z_{i})$ is the unit vector in the direction $\Omega_{i}$. 
However, for the product state $(\ket{0}_1\ket{0}_2$) we see a distinction in angular dependence:
\begin{eqnarray}
W_{\SU{2^{[2]}}} & = & \frac14 \left(1+\sqrt 5 (z_{A}+z_{B}) + \sqrt 5 z_{A}z_{B}\right) \\
W_{\bigotimes^2 \SU{2}} & = & \frac14 \left(1+\sqrt 3 (z_{A}+z_{B}) +3z_{A}z_{B}\right).
\end{eqnarray}
Note that the one-qubit and two-qubit angular terms carry coefficients of different magnitude in the tensor-product Wigner function. 

The above distinctions have led us to speculate that the two different forms of the Wigner function that we consider in this paper may be useful as a mechanism to differentiate (in representation) logical and physical qubit systems. 
That is, when there is a natural separation into physical qubits, into subsystems, or into a system and an environment, we choose the tensor product formulation. 
If, on the other hand, the system under consideration comprises a many-level quantum system constrained to act as logical qubits, it is less natural to impose a tensor product structure to the phase space representation than use the full-group form, which may be more natural.
Furthermore, in systems that comprise a mixture of logical and physical qubits a tensor product of the different kernels could be used to maintain this distinction. 
We also note that in the case of $W_{\bigotimes^2 \SU{2}}$ the Weyl transform $\DO = \int_{\Param} \Func{\DO} \Oper \ud \Param$ reconstructs the original density matrix but in the case of $W_{\SU{2^{[2]}}}$ a further linear transform is needed. 
If reconstruction of the density matrix from the Wigner function is desired, $W_{\bigotimes^2 \SU{2}}$ would be the more appropriate choice.
While much further work needs to be done, it may well be that drawing such distinctions may help us understand separability from a phase space perspective, thus enabling derivation of new useful entanglement measures.

\section{Concluding remarks}
We have demonstrated a simple method for quantum state reconstruction that extends those previously known for quantum optical systems~\cite{PhysRevLett.78.2547,PhysRevA.60.674,PhysRevLett.89.200402,PhysRevLett.87.050402,PhysRevA.70.053821,Deleglise:2008gt} to other classes of systems. 
Using IBM's \emph{Quantum Experience} five-qubit quantum processor, we have shown reconstruction of two Bell states and the five-qubit GHZ spin Schr\"odinger cat state via spin Wigner function measurements. 
We note that our procedure could be made much more efficient by direct implementation of rotation operations and measurement of any suitable extended parity operator (or, if appropriate, direct measurement of the rotated extended parity). 
By doing so, the potential advantage of our procedure over other tomographic methods would be made much clearer in that fewer measurements would be needed to check certain properties of the quantum state.
In such an instance, in verifying the fidelity of a high-quality GHZ state, only a small set of measurements that quantifies the strength of the interference terms is needed, providing an improvement over traditional quantum state tomography.
Furthermore, this work demonstrates how phase space methods can be of utility in understanding processes such as decoherence and be useful in the ``debugging'' of quantum information processors. 
In particular we have proposed a method for verifying a system is entangled  that uses only a few measurements and which in some circumstances where suitable constraints of the range of possible states are known may potentially be reduced to only two.
The utility of this work extends beyond metrology as the inclusion of tomography in device engineering will no doubt be of use in the development of quantum analogues for ``Design for Test'',  debug, fault identification and system certification.

\begin{acknowledgments}
We would like to thank Jay Gambetta, Lev Bishop, Ray Bishop, and Simon Devitt for interesting and informative discussions. 
We are deeply grateful to IBM's Quantum Computing research team and the IBM \emph{Quantum Experience} project which made it possible for us to easily obtain experimental results (\url{http://www.research.ibm.com/quantum/}). 
The views expressed are those of the authors and do not reflect the views or opinions of IBM or any of its employees.
\end{acknowledgments}

\textbf{Supplementary material: }  We provide three supporting animations, showing the evolution of $\mathrm{W}(\Omega_A,\Omega_B)=W_{\SU{2^{[2]}}}(\Omega_A,\Omega_B)$ (animations 1 and 2) or $\mathrm{W}(\Omega_A,\Omega_B)=W_{\bigotimes^2 \SU{2}}(\Omega_A,\Omega_B)$ (animation 3), where $\Omega_i=(\theta_i,\varphi_i)$.  
The Bloch spheres in the top left show four two-dimensional slices $\mathrm{W}(\Omega,R(\Omega))$, where $R$ represents the identity or a $\pi$ rotation about each of the three coordinate axes.  
The plots in the top right show the Wigner function as a function of $\theta_A$ and $\theta_B$ for fixed values of $\varphi_A$ and $\varphi_B$. 
The bottom right panel shows the Wigner functions for the individual qubits calculated from their reduced density matrices. The bottom left panel shows the progress of the simulation through the algorithm and the entanglement entropy.

In \texttt{animation\_01} we show the $W_{\SU{2^{[2]}}}$-Wigner function dynamics for the Deutsch algorithm for two qubits where the $U_f$ gate is a C-NOT gate.  
Note that there is no entanglement and the maximum value of the Wigner function for the individual qubits corresponds to the equivalent point on the Bloch-sphere. 
In \texttt{animation\_02} we show the $W_{\SU{2^{[2]}}}$-Wigner function dynamics for the creation of the four Bell states. Here we see that the Wigner functions for these states are rotations of each other in four dimensional space (indeed this is true for any maximally entangled state of two qubits). 
In \texttt{animation\_03} we show a $W_{\bigotimes^2 \SU{2}}$-Wigner function version of \texttt{animation\_02} for comparative purposes.


%
%
\bibliographystyle{apsrev4-1}
\bibliography{refs}  

\end{document}